\documentclass[review,fleqn]{elsarticle}
\usepackage[paperheight=12in,paperwidth=10.5in]{geometry}
\pdfoutput=1
\usepackage{graphicx}
\usepackage{adjustbox, lipsum}
\usepackage{colortbl}
\usepackage{amssymb}
\usepackage{amsmath}
\newenvironment{rcases}
  {\left.\begin{aligned}}
  {\end{aligned}\right\rbrace}
\usepackage{hyperref}
\journal{Physica A}
\begin{document}

\begin{frontmatter}

\title{Moran-evolution of cooperation: From well-mixed to heterogeneous complex networks}

\author{Bijan Sarkar}
\address{Department of Mathematics, Neotia Institute of Technology, Management and Science,\\
              Diamond Harbour Road, 24 Parganas (South), \\
              West Bengal-743368, India.\\}

\ead{bijan0317@yahoo.com; bijan0317@gmail.com}

\begin{abstract}
Configurational arrangement of network architecture and interaction character of individuals are two most influential factors on the mechanisms underlying the evolutionary outcome of cooperation, which is explained by the well-established framework of evolutionary game theory. In the current study, not only qualitatively but also quantitatively, we measure Moran-evolution of cooperation to support an analytical agreement based on the consequences of the replicator equation in a finite population. The validity of the measurement has been double-checked in the well-mixed network by the Langevin stochastic differential equation and the Gillespie-algorithmic version of Moran-evolution, while in a structured network, the measurement of accuracy is verified by the standard numerical simulation. Considering the Birth-Death and Death-Birth updating rules through diffusion of individuals, the investigation is carried out in the wide range of game environments those relate to the various social dilemmas where we are able to draw a new rigorous mathematical track to tackle the heterogeneity of complex networks. The set of modified criteria reveals the exact fact about the emergence and maintenance of cooperation in the structured population. We find that in general, nature promotes the environment of coexistent traits. 
\end{abstract}

\begin{keyword}
Analytical measurement \sep Stochastic evolutionary game dynamics \sep Pair approximation \sep BD and DB updating rules \sep Graph heterogeneity

\end{keyword}

\end{frontmatter}

\section{Introduction}
\label{intro}
The simple stochastic process, Moran process has long been known in biology to describe the outcome of an evolutionary process in a finite population in which two distinct alleles are competing for dominance. In particular, evolution on networks, the frequency dependent Moran-evolution moves ahead through two popular update rules: Birth-Death updating and Death-Birth updating \citep{nowak2004emergence, ohtsuki2006simple, taylor2007evolution, zukewich2013consolidating}. Albeit the order of the two events in each rule is not same, both of the rules follow the same methodology to choose an individual for reproduction, and the choice is  defined by a probability which is proportional to its fitness while an individual is chosen randomly for death. The phenomena corresponding to the particular social dilemmas are incorporated with the update rules in order to measure the evolution of cooperation under a convenient framework  of evolutionary game theory for the modelling of the well-mixed as well as structured populations.

A well-mixed population can be described as a structured population where in each time step each of the individuals has an opportunity to interact with every other individual of the population, i.e., in network language a well-mixed network is a complete graph with having the variance of the degree distribution equal to zero in respect of time as well as position. Traditionally, frequency dependent evolutionary dynamics have been studied on infinitely large homogeneous population. Around the middle of past decade the initiative was taken to combine the evolutionary game theory and the Moran process in order to investigate the evolutionary dynamics in finite population \citep{nowak2004emergence,taylor2004evolutionary}. On the level of individuals, identified three microscopic stochastic features viz.: selection, reproduction and replacement, Traulsen et al. \cite{traulsen2005coevolutionary} successfully derived the standard and the adjusted replicator dynamics. It has recently been reported that the aspiration dynamics in a structured population and the evolutionary dynamics in the corresponding well-mixed population follow  nearly the same evolutionary outcome path \citep{du2015aspiration}.

Since  the cooperative behaviour is very natural to observe in various forms of life system of: bacteria, plants, animal, human society, tissue architecture in multi-cellular organism, etc., how cooperation emerging and being enhanced are the most captivating section of evolutionary dynamics. For a structured population, the limited dispersal act in individuals ensures that the offspring of a parent is located close to  the parent, i.e., the act enables cooperators to form a cluster for which the boosting capacity to increase the frequency interaction rate in cooperator individuals is higher than that of in random interaction. Such assortment between cooperators and defectors is one of the explanations of the enhancement of cooperation in a structured pattern \citep{fletcher2009simple, perc2013evolutionary}. However, Perc \citep{perc2011does} shows that the expected promotion of cooperation may be less for games governed by group interactions than originally assumed.

For representing a population structure, we use the mathematical tool of evolutionary graph theory: vertices correspond to individuals and edges indicate interactions \citep{szabo2007evolutionary}. Using pair approximation (see for application procedure: \citep{matsuda1992statistical, morris1997representing, van1998unit, house2011insights, hadjichrysanthou2012approximating}) in the limit of weak selection, the analytical theory about the emergence and the enhancement of cooperation on a regular structure was developed where the theory itself revels that benefit-to-cost ratio is to be greater than the degree of the graph \citep{ohtsuki2006simple} -- or greater than the mean degree of the nearest neighbours \citep{konno2011condition}. Later, the condition straight forwardly is utilised to derive the replicator equation on graph \citep{ohtsuki2006replicator}. As in this assumption of evolutionary process there is no mutation just selection, the offspring of each individual is a perfect copy of its parent. It is generally believed that the graph heterogeneity dramatically enhances cooperation in complex network \citep{santos2006evolutionary, perc2008social, santos2012role, xu2015evolution} where for accounting a range of heterogeneities at the level of individuals, networks theory provides a new modelling paradigm. In the past few years, from multiple viewpoints the effect of nontrivial topologies on evolution of cooperation has been explored \citep{perc2009evolution, szolnoki2016collective, perc2017statistical}. Analytically, evolutionary games on heterogeneous graphs have mainly investigated under Death-Birth updating and Pair-Wise Comparison updating \citep{santos2008social,  szolnoki2008towards, pinheiro2012local}. Based on theoretical study, Li et al. \cite{li2013evolution} recently introduces a new  state variable to measure the time evolution of cooperation on a heterogeneous graph. This is the first model in which rigorous mathematical procedure is followed. Besides the network reciprocity, in the theory of how and why cooperation emerges the effect of individuals' movement through the form of diffusion can also be counted, as a direct reciprocity, easily \citep{nowak2006five}.

Frequently, in material and natural world when the population size is relatively small, each member of the population generally interacts with any other member of the population in a random way. Our intuitive sense also suggests that if well-mixed and pattern interactions run parallel, the well-mixed prevails over the pattern interaction. Additionally, we know that world wild life follows the random interaction, their evolution has strong influence to control the eco-system. As we may use the update rules here too, replicator equation, a measurement procedure of evolution inevitably plays a crucial role to measure the whole scenarios.

In the present article, $p(k,\tilde{k})$ denotes the conditional probability of a vertex with degree $k$ connected to other vertices with degree $\tilde{k}$. We consider that for vertices with degree $k$, degree distribution of a graph is equal to $p(k)$. That is, the considered vertex has $k$-number of links. The probability that each of these links is attached to a vertex with degree $\tilde{k}$ is given by: $kp(\tilde{k})$. More precisely, the probability that an end of a link is attached to a vertex with degree $\tilde{k}$ is given by: $\frac{kp(\tilde{k})}{z}$, $z$ being average degree of the graph. Therefore, we have:  $p(k,\tilde{k})= p(k). \frac{kp(\tilde{k})}{z}$. In the model of Moran-evolution, $p(k, \tilde{k})$ -- a non-symmetric function -- is to be implicitly involved with the transition probabilities.

In the present treatment, the analytical measurement tool of Moran-evolution of cooperation in the well-mixed population as well as in the structured population is derived as a form of replicator equation to check the accuracy of predictions about the outcomes of an evolution. The accuracy is evaluated through stochastic simulation in the comparison mode. Considering the Birth-Death and Death-Birth processes on homogeneous and heterogeneous networks in the wide range of game environments, every possible outcome of evolution has been attained and focused for a logical explanation. The effects on the outcomes also take into account the diffusion of individuals on networks. We point out many new results along with the known results. Here, modeller's main intention is that find out the right path so that we are able to determine the appropriate ranges of the influential factors of Moran-evolution, properly.

\section{The well-mixed network effect on Moran-evolution}
\label{sec:1}

In order to investigate the  Moran-evolution in finite population, we consider the evolution of a population with two strategies $C$ and $D$. The state variable $i$ denotes the number of $C$ players and $N-i$ $D$ players where the population size $N$ is constant as in each discrete time step one birth and one death occur. Fitness of $C$ and $D$, respectively, is given by: $f_{i}=1-w+wF_{i}$, $g_{i}=1-w+wG_{i}$. The parameter $w$ that measures the intensity of the selections is a number between $0$ and $1$. Strategies $C$ and $D$ are neutral variants if $w=0$. If $w=1$ selection is strong; the fitness is entirely determined by the expected payoffs $F_{i}$ and $G_{i}$. And $w \ll 1$ interprets the case of weak selection where the payoff contributes only a small effect on fitness. Every individual interacts with each representative individual of the considering population where the expected payoff of $C$ and $D$ individuals is determined by the fraction of coplayers of both types. If the payoffs of an $i^{\mbox{th}}$ individual getting from each interaction with a $j^{\mbox{th}}$ individual are denoted by the $\pi_{ij}$, then excluding self-interactions, the expected payoffs are given by 
\begin{eqnarray}
F_{i}=\frac{\pi_{CC}(i-1)+\pi_{CD}(N-i)}{N-1}, \hspace{5mm}  G_{i}=\frac{\pi_{DC}i+\pi_{DD}(N-i-1)}{N-1}. 
\end{eqnarray}

In the Moran process, the evolution is defined by the transition probabilities $T^{+}_{i}$: moving $i$ to $i+1$ $C$ players , $T^{-}_{i}$: moving $i$ to $i-1$ $C$ players and the process remains in state $i$ is simply $T^{0}_{i}=1-T^{+}_{i}-T^{-}_{i}$. All other transitions have zero probabilities. Introducing the coefficient $\theta_{i}$ that captures the effects of population structure and update rules in weak selection we have, $\frac{T_{i}^{-}}{T_{i}^{+}}\approx 1+w\theta_{i}$. The  mixed population of $C$ and $D$ will eventually end up in either all$-C$ or all$-D$ through this process, and the occurrence is measured by the fixation probability. Here, fixation probabilities  of $C$ and $D$ respectively can be calculated as:
\begin{eqnarray}
\rho_{C}=\frac{1}{1+\sum_{j=1}^{N-1}\prod_{i=1}^{j}(\frac{T_{i}^{-}}{T_{i}^{+}})}\approx \frac{1}{N}-\frac{w}{N^{2}}\sum_{i=1}^{N-i}(N-i)\theta_{i},\\
\rho_{D}=\frac{1}{1+\sum_{j=1}^{N-1}\prod_{i=1}^{j}(\frac{T_{N-i}^{+}}{T_{N-i}^{-}})}\approx \frac{1}{N}+\frac{w}{N^{2}}\sum_{i=1}^{N-i}(N-i)\theta_{N-i}.
\end{eqnarray}
\noindent This is the one of the procedures to deal with the evolutionary outcomes of cooperation where is to be the main task to  calculate the values of fixation probabilities. Zukewich et al. \cite{zukewich2013consolidating} shows that the coefficient $\theta_{i}$ plays the same role as the fixation probability; thus we can avoid the diffusion approximation procedure \citep{kimura1964diffusion}. However, in the present  study, we do not follow this track because we are interested to derive a deterministic dynamical system of cooperation in weak selection. Here, we also note that if $x_{i}$ is the fraction of chance of ending up the process in state $N$ when starting from state $i$, then at the neutral drift (i.e., where $f_i=g_i=1$) we have $x_{i}=\frac{i}{N}$.

Under the Birth-Death (BD) procedure in the case of well-mixed population, the transition probability of the number of $C$'s increasing from $i$ to $i+1$ when a $C$ reproduces and a $D$ dies is
\begin{eqnarray}
T_{i}^{+}&=&\frac{if_{i}}{if_{i}+(N-i)g_{i}} \cdot \frac{N-i}{N} \nonumber \\
         &=& \frac{i}{N}\frac{N-i}{N}+\omega \frac{i}{N}(\frac{N-i}{N})^2(F_{i}-G_{i})+\mathcal{O}(\omega^2) \nonumber \\
         &=& x_{i}(1-x_{i})[1+\frac{\omega (1-x_{i})}{1-\frac{1}{N}}\{ (\pi_{CC}-\pi_{DC})x_{i} \nonumber \\ 
         & &+(\pi_{CD}-\pi_{DD})(1-x_{i})-\frac{1}{N}(\pi_{CC}-\pi_{DD})\}] +\mathcal{O}(\omega^2), 
\end{eqnarray}
and the transition probability of the number of $C$'s decreasing from $i$ to $i-1$ when a $D$ reproduces and a $C$ dies is
\begin{eqnarray}
T_{i}^{-}&=&\frac{(N-i)g_{i}}{if_{i}+(N-i)g_{i}} \cdot \frac{i}{N} \nonumber \\
         &=& \frac{i}{N}\frac{N-i}{N}+\omega (\frac{i}{N})^2\frac{N-i}{N}(G_{i}-F_{i})+\mathcal{O}(\omega^2) \nonumber \\
         &=& x_{i}(1-x_{i})[1-\frac{\omega x_{i}}{1-\frac{1}{N}} \{ (\pi_{CC}-\pi_{DC})x_{i} \nonumber \\ 
         & &+(\pi_{CD}-\pi_{DD})(1-x_{i})-\frac{1}{N}(\pi_{CC}-\pi_{DD})\}] +\mathcal{O}(\omega^2). 
\end{eqnarray}
\noindent In similar fashion, we can calculate the transition probabilities of Moran-evolution under the Death-Birth (DB) process. However, for $N \rightarrow \infty $, the transition probabilities are same under the  BD  and  DB processes, i.e., both the updates lead to the same dynamics. Thus, here is no need to look at the DB version of Moran-evolution in the well-mixed population.

Generally, the evolution of the two strategies $C$ and $D$ is described by the deterministic replicator like dynamics for large populations and the stochastic evolutionary game dynamics in finite populations. And, both dynamics are related to each other through the following mathematical formalism. Introducing the notations $x=\lim_{N \to \infty}x_{i}$, $t=\frac{\tau}{N}$, setting the probability density $c(x_{i},t)=NP^{\tau}(i)$ where $P^{\tau}(i)$ is the probability that the system is in the state $i$ at time $\tau$, and replacing the simple rate laws by probability laws, we logically get
\begin{eqnarray}
\hspace{-4mm}c(x_{i}, t+\frac{1}{N})-c(x_{i},t)&=&c(x_{i}-\frac{1}{N},t)T^{+}(x_{i}-\frac{1}{N})+c(x_{i}+\frac{1}{N},t)T^{-}(x_{i}+\frac{1}{N})\nonumber \\
                          & & -c(x_{i},t)T^{+}(x_{i})-c(x_{i},t)T^{-}(x_{i}). \nonumber                    
\end{eqnarray} 

In the limit of large $N$, the probability densities and the transition probabilities are expanded in a Taylor series at $x_{i}$ and $t$ upto second order in $N^{-1}$; the step yields
\begin{eqnarray}
\frac{\partial c}{\partial t}=- \frac{\partial }{\partial x_{i}}[a(x_{i})c(x_{i},t)]+\frac{1}{2}\frac{\partial^2 }{\partial x_{i}^2}[b^{2}(x_{i})c(x_{i},t)],
\end{eqnarray} 
\noindent where $a(x_{i})=T^{+}(x_{i})-T^{-}(x_{i})$ and $b(x_{i})=\sqrt{\frac{1}{N}[T^{+}(x_{i})+T^{-}(x_{i})]}$. This equation is nothing but the form of  Fokker-Plank equation for finite and large value of $N$. Now, we know that white noise process $\xi(t)$ is formally defined as the derivative of the Brownian motion:
\[\xi(t)=\frac{dB(t)}{dt}=B^{\prime}(t), \] 
\noindent and as $a(x_{i})$ and $b(x_{i})$ are independent of $\xi(t)$, using the Ito calculus, we can derive a Langevin equation
\begin{eqnarray}
\dot{x_{i}}=a(x_{i})+b(x_{i})\xi(t).
\end{eqnarray} 
\noindent We know that the act of  Ito's formula can be expressed as \citep{Kampen1992, Gardiner2009}
\begin{eqnarray}
<\frac{df[x(t)]}{dt}>&=&\frac{d}{dt}<f[x(t)]> \nonumber \\
                     &=& <a[x(t)]\partial_{x}f+\frac{1}{2}b^{2}[x(t)]\partial^{2}_{x}f>. \nonumber
\end{eqnarray}
\noindent That is 
\begin{eqnarray}
\int{dx f(x)\partial_{t}c(x,t)}=\int{dx[a(x)\partial_{x}f+\frac{1}{2}b^{2}(x)\partial^{2}_{x}f]c(x,t)}. \nonumber
\end{eqnarray}
\noindent Next, integration by parts gives 
\[ \int{dx f(x) \partial_{t}c(x,t)}=\int{dx f(x)(-\partial_{x}[a(x)c]+\frac{1}{2}\partial^{2}_{x}[b^{2}(x)c])}+\mbox{boundary terms}.\] 
\noindent Under the considering environment, the boundary terms will be vanished and the arbitrariness of $f(x)$ reveals the fact that the diffusion process could be locally approximated by a Langevin equation.

For relatively large value of $N$, the noise term, $b(x)\xi(t)$,  disappears  and  only the term $a(x)$ determines the dynamics. This case reduces to the following deterministic differential equation
\begin{eqnarray}
\dot{x}=T^{+}(x)-T^{-}(x), \label{eq:8}
\end{eqnarray}
\noindent and recovers the replicator like equation. The exact form of this equation is written as :
\begin{eqnarray}
\frac{dx}{dt} = \omega x(1-x)[\pi_{CC}x+\pi_{CD}(1-x)-\pi_{DC}x-\pi_{DD}(1-x)]+\mathcal{O}(\omega^2), \label{eq:9}
\end{eqnarray}
\noindent where the constant factor $\omega$ influences on the time scale. It is clear that for the using of the unmodified expressions of transition probabilities related to Moran process, the Eq.(\ref{eq:8}) turns out to be the third order standard replicator equation instead of the adjusted replicator equation (compare to \citep{traulsen2005coevolutionary}). Since Eq.(\ref{eq:9}) does not depend on the probability terms explicitly even though its derivation depends on the classical probability concept, here we consider the equation as  a deterministic one.

Next, we have to take an initiative to construct a stochastic platform to measure the accuracy of the outcomes of the deterministic model. Here, the stochastic version of Moran-evolution of cooperation in finite populations can be portrayed through the Langevin stochastic differential equation as well as Gillespie algorithm \citep{gillespie1976general,gillespie1977exact}, where Moran-evolution is defined by Gillespie algorithm in the form of the two reactions: Cooperator+ Defector $\rightarrow$ Cooperator+ Cooperator, and Cooperator+Defector $\rightarrow$ Defector + Defector. In this section, we confine ourself to derive the governing equation of evolution of $C$ (equivalently or $D$) individual which we have already obtained. Later, in the subsection \ref{subsec:4.1} we numerically show that on what extend the Moran-evolution of cooperation in the analytical method and in the stochastic method are agreed  with each other in the flavours of various game environments; those are defined by the set of two game parameters $(u, v)$. For the comparison with the discrete evolution graph model in the finite population, the well-mixed results have also be taken into consideration in the subsection \ref{subsec:4.2}.

\section{The heterogeneous complex network effect on Moran-evolution}
\label{sec:3}
\subsection{Birth-Death updating through  diffusion of individuals}
\label{subsec:3.1}

We look attentively at a Moran process on a directed graph with $N$ vertices and degree distribution $p(k)$ for $k=1,2,......,N-1$. The individual at each vertex $i$ uses either strategy $C$ or strategy $D$ in a mutual and reciprocal action with all of its neighbours where the degree of vertex is denoted by $k_{i}$ for $i=1,2,......,N$. Thus, the average degree of the graph is $z=\sum_{i=1}^{N}k_{i}/N$ or $z=\sum_{k}kp(k)$ which is the expected number of neighbours of an individual chosen  at random. As the total  number of directed edges in the graph is $zN$, the proportions of the directed  edges starting from cooperators and defectors, denoted by $\phi_{C}$  and $\phi_{D}$, respectively, are defined as $\phi_{C}=\sum_{s_{i}=C}k_{i}/zN$ and $\phi_{D}=\sum_{s_{i}=D}k_{i}/zN$, where $s_{i}$ is the strategy of the individual at the vertex $i$ i.e., $s_{i}\in \{C,D\}$ for $i=1,2,......,N$. By the using of the observables $\phi_{k,C}$  and $\phi_{k,D}$ --  the proportion of vertices with degree $k$ those are in strategy $C$ and in strategy $D$, respectively, --  the $\phi_{C}$ and $\phi_{D}$ can also be expressed as  $\phi_{C}=\frac{1}{z} \sum_{k}kp(k)\phi_{k,C}$ and $\phi_{D}=\frac{1}{z} \sum_{k}kp(k)\phi_{k,D}$. In this article, $\phi_{C}$ $( \phi_{D} )$ is called the frequency of $C$ individuals ($D$ individuals) in the structured framework of the finite population. It is to note that for the regular graph, i.e., $k_{i}=z$ for all $i=1,2,......,N$, the $\phi_{C}$ is equal to $x$ -- the frequency of $C$ individuals in the well-mixed population (see \citep{li2013evolution}).

For $X,Y \in \{C,D\}$, $\Phi_{XY}$ denotes the number of directed edges $(i,j)$ with $s_{i}=X$ and $s_{j}=Y$, that is the proportion of directed edges with strategy pair $(X,Y)$ in the total  set of directed edges is defined as $\phi_{XY}=\frac{\Phi_{XY}}{\Phi_{CC}+\Phi_{CD}+\Phi_{DC}+\Phi_{DD}}=\Phi_{XY}/zN$. And under the pair approximation, the conditional probability that a neighbour of a vertex with strategy $Y$ is a $X-$individual is given by $q_{X|Y}=\phi_{YX}/\phi_{Y}$. Here, evolutionary dynamics on a heterogeneous graph is described by the variables $\phi_{X}$, $q_{X|Y}$ and $\phi_{XY}$ where $\phi_{C}+\phi_{D}=1$, $q_{C|X}+q_{D|X}=1$, $\phi_{YX}= q_{X|Y}\phi_{Y}$ and $\phi_{CD}=\phi_{DC}$. More precisely, we can show that the entire system can be drawn by only two variables $\phi_{C}$ and $\phi_{CC}$.

To continue the flow, we first take Birth-Death (BD) updating process  as one of our main intentions is to derive a rigorous mathematical formulation to measure the Moran-evolution of cooperation under BD updating on a heterogeneous complex network in a finite population.  For BD updating an individual is randomly chosen to reproduce with a probability proportional to fitness and its offspring then replaces a random neighbour. If a cooperator with degree $k$ is randomly chosen to reproduce, then  according to binomial theorem the probability that this focal individual has exactly $k_{C}$ neighbours with strategy $C$ and $k_{D}$ neighbours with strategy $D$ is $\frac{k!}{k_{C}!k_{D}!}(q_{C|C})^{k_{C}}(q_{D|C})^{k_{D}}$ and it would have a fitness of: $f_{0}=1-\omega+\omega[\frac{k_{C}}{k}\pi_{CC}+\frac{k_{D}}{k}\pi_{CD}]$. Similar to previous, the payoffs of an $i^{\mbox{th}}$ individual getting from each interaction with a ${j^{\mbox{th}}}$ individual are denoted by the $\pi_{ij}$. On the network, if the other-end individual of the considering link having focal $C-$ individual, is of $D-$individual with $\tilde{k}$ degree, then the probability of existence of such configuration would logically be taken as $p(k,\tilde{k})=p(k).\frac{kp(\tilde{k})}{z}$ in respect of the focal individual, and in respect of the other-end individual this probability would then be given by $p(\tilde{k},k)=p(\tilde{k}).\frac{\tilde{k}p(k)}{z}$; that is  the  conditional probability function is non-symmetric, clearly. Therefore, to probability that the $\phi_{C}$ increases by $\tilde{k}/zN$ in a time step is the probability that a $C$ reproduces and a $D$ then dies, which is given by
\begin{eqnarray}
\Pr(\Delta \phi_{C}=\frac{\tilde{k}}{zN})&=&p(k)\phi_{k,C} \sum_{k_C+k_D=k}\frac{k!}{k_{C}!k_{D}!}(q_{C|C})^{k_{C}}(q_{D|C})^{k_{D}} \cdot \frac{f_{0}}{\vartheta}\cdot \frac{k-k_{C}}{k} \nonumber \\ &  &  \times
\frac{k p(\tilde{k})}{z} \cdot \sum_{\tilde{k}_C+\tilde{k}_D=\tilde{k}-1}\frac{(\tilde{k}-1)!}{\tilde{k}_{C}!\tilde{k}_{D}!}(q_{C|D})^{\tilde{k}_{C}}(q_{D|D})^{\tilde{k}_{D}},
\end{eqnarray}
\noindent and taking the normalized expression of average existence of cooperators at particular $\tilde{k}_{C}$ numbers of cooperators in $\tilde{k}$ numbers of total neighbours as: {\tiny $(\frac{\tilde{k}_{C}}{\tilde{k}})\frac{\tilde{k}!}{\tilde{k}_{C}!\tilde{k}_{D}!}(q_{C|D})^{\tilde{k}_{C}}(q_{D|D})^{\tilde{k}_{D}}$}, 
the probability that $\phi_{CC}$ increases by $2\tilde{k}_{C}/zN$ can be calculated as 

\begin{small}
\begin{eqnarray}
\Pr(\Delta \phi_{CC}=\frac{2\tilde{k}_{C}}{zN})&=&\sum_{ k_C+k_D=k-1} p(k) \cdot \frac{(k-1)!}{k_{C}!(k-k_{C}-1)!}(q_{C|C})^{k_{C}}(q_{D|C})^{k-k_{C}-1} \cdot \frac{f_{0}}{\vartheta} 
 \nonumber \\ 
 &  & \times 
 \sum_{\tilde{k} \geq \tilde{k}_C} \frac{k p(\tilde{k})}{z} \cdot \phi_{\tilde{k},D} \cdot (\frac{\tilde{k}_{C}}{\tilde{k}}) \frac{\tilde{k}!}{\tilde{k}_{C}!(\tilde{k}-\tilde{k}_{C})!}(q_{C|D})^{\tilde{k}_{C}}(q_{D|D})^{\tilde{k}-\tilde{k}_{C}},
\end{eqnarray}
\end{small}

\noindent where $\vartheta$ denotes a normalization constant, which is given by the average  fitness over all individuals in the population (for the notation interpretation see \citep{Morita2008}):
\[ \vartheta= 1-\omega +\omega [p_{C}(\pi_{CD}+(\pi_{CC}-\pi_{CD})q_{C|C})+(1-p_{C})(\pi_{DC}+(\pi_{DD}-\pi_{DC})q_{D|D})].\]
\noindent Similarly, if a defector  with degree $k$ is randomly chosen to reproduce, then  according to binomial theorem the probability that this focal individual has exactly $k_{C}$ neighbours with strategy $C$ and $k_{D}$ neighbours with strategy $D$ is $\frac{k!}{k_{C}!k_{D}!}(q_{C|D})^{k_{C}}(q_{D|D})^{k_{D}}$ and it would have a fitness of: $g_{0}=1-\omega+\omega[\frac{k_{C}}{k}\pi_{DC}+\frac{k_{D}}{k}\pi_{DD}]$.
To probability that the $\phi_{C}$ decreases by $\tilde{k}/zN$ in a time step is the probability that a $D$ reproduces and a $C$ then dies, which is given by
\begin{eqnarray}
 \Pr(\Delta \phi_{C}=-\frac{\tilde{k}}{zN})&=&p(k)\phi_{k,D} \sum_{k_C+k_D=k}\frac{k!}{k_{C}!k_{D}!}(q_{C|D})^{k_{C}}(q_{D|D})^{k_{D}} \cdot \frac{g_{0}}{\vartheta} \cdot \frac{k_{C}}{k} \nonumber \\ &  &  \times
\frac{k p(\tilde{k})}{z} \cdot \sum_{\tilde{k}_C+\tilde{k}_D=\tilde{k}-1}\frac{(\tilde{k}-1)!}{\tilde{k}_{C}!\tilde{k}_{D}!}(q_{C|C})^{\tilde{k}_{C}}(q_{D|C})^{\tilde{k}_{D}},
\end{eqnarray}
\noindent and taking the normalized expression of average existence of defectors at particular $\tilde{k}_{D}$ numbers of defectors in $\tilde{k}$ numbers of total neighbours as: {\tiny $(\frac{\tilde{k}_{D}}{\tilde{k}}) \frac{\tilde{k}!}{\tilde{k}_{C}!\tilde{k}_{D}!}(q_{C|C})^{\tilde{k}_{C}}(q_{D|C})^{\tilde{k}_{D}}$}, the probability that $\phi_{CC}$ decreases by $2\tilde{k}_{C}/zN$ can be calculated as 

\begin{small}
\begin{eqnarray}
 \Pr(\Delta \phi_{CC}= -\frac{2\tilde{k}_{C}}{zN})&=&\sum_{ k_C+k_D=k-1} p(k) \cdot \frac{(k-1)!}{k_{D}!(k-k_{D}-1)!}(q_{C|D})^{k-k_{D}-1}(q_{D|D})^{k_{D}} \cdot \frac{g_{0}}{\vartheta} \nonumber \\   
&  & \times \sum_{\tilde{k} \geq \tilde{k}_C}\frac{k p(\tilde{k})}{z}\cdot \phi_{\tilde{k},C}\cdot(\frac{\tilde{k}_{D}}{\tilde{k}}) \frac{\tilde{k}!}{\tilde{k}_{C}!(\tilde{k}-\tilde{k}_{C})!}(q_{C|C})^{\tilde{k}_{C}}(q_{D|C})^{\tilde{k}-\tilde{k}_{C}}.
\end{eqnarray}
\end{small}

The natural movement in each of the time steps by interchanging the locations of $C$ individual and $D$ individual on a network is called here diffusion of individuals where two neighbouring individuals are selected randomly. Now, we incorporate the effect of diffusion in this structured model with probability $q$, and through the diffusion process, the probability and the expectation of the rate of change of $\phi_{C}$ by $\frac{k-\hat{k}}{zN}$,  respectively, are
\begin{small}
\begin{eqnarray}
\Pr(\Delta \phi_{C}=\frac{k-\hat{k}}{zN})&=& p(k,\hat{k}),
\end{eqnarray}
\end{small}
\noindent and
\begin{small} 
\begin{eqnarray}
<(\Delta \phi_{C}=\frac{k-\hat{k}}{zN})>&=& \sum_{\hat{k}=1}^{N-1}\sum_{k=1}^{N-1} \frac{k-\hat{k}}{zN}\Pr(\Delta \phi_{C}=\frac{k-\hat{k}}{zN}) \nonumber \\
                                       &=& 0.
\end{eqnarray}
\end{small}

\noindent The simple conclusion is: diffusion does not directly have any effect on the concentration of $\phi$. In this context, the probabilities related to the two links that $\phi_{CC}$ increases or decreases by $2 (k_{C}-\hat{k}_{C})/zN$ are obtained as

\begin{small}
\begin{eqnarray}
 \Pr(\Delta \phi_{CC}= \frac{2(\hat{k}_{C}-k_{C})}{zN})&=& \phi_{CD} p(k,\hat{k})
        \sum_{k \geq k_C} \frac{(k-1)!}{k_{C}!(k-k_{C}-1)!}(q_{C|C})^{k_{C}}(q_{D|C})^{k-k_{C}-1}\nonumber \\
       & & \times \sum_{\hat{k} \geq \hat{k}_C} \frac{(\hat{k}-1)!}{\hat{k}_{C}!(\hat{k}-\hat{k}_{C}-1)!}(q_{C|D})^{\hat{k}_{C}}(q_{D|D})^{\hat{k}-\hat{k}_{C}-1},
\end{eqnarray}
\end{small}

\begin{small}
\begin{eqnarray}
 \Pr(\Delta \phi_{CC}= \frac{2(k_{C}-\hat{k}_{C})}{zN})&=& \phi_{DC} p(k,\hat{k})
        \sum_{k \geq k_C} \frac{(k-1)!}{k_{C}!(k-k_{C}-1)!}(q_{C|D})^{k_{C}}(q_{D|D})^{k-k_{C}-1}\nonumber \\
       & & \times \sum_{\hat{k} \geq \hat{k}_C} \frac{(\hat{k}-1)!}{\hat{k}_{C}!(\hat{k}-\hat{k}_{C}-1)!}(q_{C|C})^{\hat{k}_{C}}(q_{D|C})^{\hat{k}-\hat{k}_{C}-1}.
\end{eqnarray}
\end{small}

\noindent That is, due to diffusion in the $(D,C)$-link -- corresponding to the degree coordinate $(k, \hat{k})$ -- the probability of average increment of $(C,C)$-link at the $D$-end is

\begin{small}
\begin{eqnarray}
\Pr^\prime(\Delta \phi_{CC}= \frac{2(k-1)}{zN}\cdot q_{C|D})&=& \phi_{DC} p(\hat{k},k),
\end{eqnarray}
\end{small}
\noindent and the probability of average decrement of $(C,C)$-link at the $C$-end is
\begin{small}
\begin{eqnarray}
\Pr^\prime(\Delta \phi_{CC}= -\frac{2(\hat{k}-1)}{zN}\cdot q_{C|C})&=& \phi_{DC} p(k,\hat{k}).
\end{eqnarray}
\end{small}
  
\noindent As the model is an asynchronous model -- like voter model, i.e., in each time step diffusion will possibly occur either in the $(C,D)$-link or in the $(D,C)$-link,  thus the time evolution of $\phi_{C}$  and $\phi_{CC}$, respectively, is given by 
\begin{eqnarray}
\frac{d\phi_{C}}{dt} = \sum_{\tilde{k}=1}^{N-1}\sum_{k=1}^{N-1}\frac{\tilde{k}}{zN}\Pr(\Delta \phi_{C}=\frac{\tilde{k}}{zN}) - \sum_{\tilde{k}=1}^{N-1}\sum_{k=1}^{N-1}\frac{\tilde{k}}{zN}\Pr (\Delta \phi_{C}=-\frac{\tilde{k}}{zN}), \label{eq:14}
\end{eqnarray}

\begin{small}
\begin{eqnarray}
\frac{d\phi_{CC}}{dt} = (1-q)&[ & \sum_{\tilde{k}_{C}=1}^{N-1}\sum_{k=1}^{N-1}\frac{2\tilde{k}_{C}}{zN}\Pr(\Delta \phi_{CC}=\frac{2\tilde{k}_{C}}{zN})-\sum_{\tilde{k}_{C}=1}^{N-1}\sum_{k=1}^{N-1}\frac{2\tilde{k}_{C}}{zN}\Pr (\Delta \phi_{CC}=-\frac{2\tilde{k}_{C}}{zN} )] \nonumber \\ 
  + q &[ & \sum_{\hat{k}=1}^{N-1}\sum_{k=1}^{N-1}\frac{2(k-1)}{zN}\cdot q_{C|D} \Pr^\prime(\Delta \phi_{CC}= \frac{2(k-1)}{zN}\cdot q_{C|D}) \nonumber \\ 
& - &\sum_{\hat{k}=1}^{N-1}\sum_{k=1}^{N-1}\frac{2(\hat{k}-1)}{zN}\cdot q_{C|C} \Pr^\prime(\Delta \phi_{CC}= -\frac{2(\hat{k}-1)}{zN}\cdot q_{C|C})]. \label{eq:15}
\end{eqnarray}
\end{small} 

It is not too hard to show that in the quasi-steady-state $\phi_{C}=\phi_{k,C}$ and $\phi_{D}=\phi_{k,D}$. We apply these relations in order to simplify the Eq.(\ref{eq:15}) and after some algebras, we have 
\begin{small}
\begin{eqnarray}
\frac{d\phi_{CC}}{dt}&=& (1-q).\frac{2\phi_{CD}}{zN}[1-(z-1)(q_{C|C}-q_{C|D})]+ q.\frac{2\phi_{CD}}{zN}[(z-1)(q_{C|D}-q_{C|C})]
                      + \mathcal{O}(\omega).
\end{eqnarray}
\end{small}

\noindent We also note that for each degree $k$, the time evolution of $\phi_{k,C}$ with consideration of diffusion process is given by
\begin{small}
\begin{eqnarray}
\frac{d\phi_{k,C}}{dt}&=& (1-q)[\frac{1}{Np(k)}\Pr(\Delta \phi_{k,C}=\frac{1}{Np(k)})-\frac{1}{Np(k)}\Pr(\Delta \phi_{k,C}=-\frac{1}{Np(k)})] \nonumber \\ 
                      & & +q[\frac{1}{Np(k)}\Pr(\Delta \phi_{k,C}=\frac{1}{Np(k)})-\frac{1}{Np(k)}\Pr(\Delta \phi_{k,C}=-\frac{1}{Np(k)})] \nonumber \\ 
                      &=& \frac{1}{Np(k)}(p(k))\phi_{k,D}q_{C|D}-p(k))\phi_{k,C}q_{D|C})+ \mathcal{O}(\omega).
\end{eqnarray}
\end{small}

As $\phi_{CC}$, $\phi_{k,C}$, $\phi_{k,D}$ are faster variables than $\phi_{C}$, those go to their quasi-steady-state while the slow variable $\phi_{C}$ stays approximately constant, and as we are interested in dynamics of $\phi_{C}$, to get the specific relations in the quasi-steady-state of the fast variables  we set $\frac{d\phi_{CC}}{dt}=0$  and $\frac{d\phi_{k,C}}{dt}=0$, the obtained relations respectively are: $(q_{C|C}-q_{C|D})=\frac{1-q}{z-1}$ and $\phi_{C}=\phi_{k,C}$. With similar fashion $\phi_{k,D}$ can be approximated by $\phi_{D}$. Therefore, the simplified form of the Eq.(\ref{eq:14}) is written as 
\begin{eqnarray}
\frac{d\phi_{C}}{dt} & = & \frac{\omega}{zN\vartheta}(\frac{z+q-2}{z-1})\phi_{C}(1-\phi_{C})[\alpha+z(\pi_{CD}-\pi_{DD})  +q(\pi_{CD}-\pi_{CC}) \nonumber \\
& & +\alpha(z+q-2)\phi_{C}]+\mathcal{O}(\omega^2),
\end{eqnarray}
where $\alpha=\pi_{CC}-\pi_{CD}-\pi_{DC}+\pi_{DD}$, and the influential time-factor is equal to $\frac{\omega}{zN\vartheta}(\frac{z+q-2}{z-1})$. Taking $t^{\prime}=\frac{\omega}{zN\vartheta}(\frac{z+q-2}{z-1})t$ and using the transition probabilities related to Moran process, the modeling formalism brings out the third-order standard replicator equation with  the transformed payoff matrix:
\begin{center}
$\bordermatrix{~ & C & D \cr
                  C & (z-1)(\pi_{CC}-\pi_{DC}) & (z-1)(\pi_{CD}-\pi_{DD}) \cr
                  D & (q-1)(\pi_{DC}-\pi_{DD}) & (q-1)(\pi_{CC}-\pi_{CD}) \cr}$. 
                  
\end{center}
\noindent In the latter section through an analysis of the three equilibria of the global dynamics of $\phi_{C}$ we mainly emphasis on how an internal equilibrium point is to be affected by the degree of a graph -- the number of links among the individuals. Well known procedure -- the perturbation method in linear stability analysis -- of ordinary differential equation reminds us that the internal equilibrium is stable when $\alpha<0$ and unstable when $\alpha>0$.

\subsection{Death-Birth updating through  diffusion of individuals}
\label{subsec:3.2}
In this update procedure in each time step an individual is chosen for death and to fill the vacant site the procedure chooses an offspring among the neighbours of the focal individual with the probability proportional to the fitness of the parent of the chosen offspring. Adopting the previous configuration structure, we get that if the randomly chosen focal individual is a defector, then the expected payoffs of its neighbours of $C$ strategist  and $D$ strategist  respectively are 
$h_{C}=(z-1)q_{C|C}(\pi_{CC}-\pi_{CD})+z\pi_{CD}$ and $h_{D}=(z-1)q_{C|D}(\pi_{DC}-\pi_{DD})+z\pi_{DD}$, 
and if the randomly chosen focal individual is a cooperator, then the expected payoff of its neighbours of $C$ strategist  and $D$ strategist  is 
$l_{C}=(z-1)q_{C|C}(\pi_{CC}-\pi_{CD})+(z-1)\pi_{CD}+\pi_{CC}$   and  
$l_{D}=(z-1)q_{C|D}(\pi_{DC}-\pi_{DD})+(z-1)\pi_{DD}+\pi_{DC}$, respectively; where for the focal individual having the defector character and the cooperator character the fitnesses of neighbour individuals are denoted by small letters $f$ and $g$ respectively with the appropriate subscripts (see the section \ref{sec:1}).

According to the update rule for the chosen defector individual with degree $k$ in which $k_{C}$ neighbors are with strategy $C$ and $k_{D}$ neighbours with strategy $D$, the probability that $\phi_{C}$ increases by $\Delta \phi_{C}=\frac{k}{zN}$ in each time step is  
\begin{small}
\begin{eqnarray}
\Pr(\Delta \phi_{C}=\frac{k}{zN})&=&p(k)\phi_{k,D} 
                   \sum_{k_C+k_D=k}\frac{k!}{k_{C}!k_{D}!}(q_{C|D})^{k_{C}}(q_{D|D})^{k_{D}} \cdot \frac{k_{C}f_{C}}{k_{C}f_{C}+k_{D}f_{D}}.
\end{eqnarray}
\end{small}
\noindent And if the chosen individual for death is cooperator, then the probability that $\phi_{C}$ decreases by $-\Delta \phi_{C}= \frac{k}{zN}$ in each time step is 
\begin{small}
\begin{eqnarray}
\Pr(\Delta \phi_{C}=-\frac{k}{zN})&=&p(k)\phi_{k,C} 
                   \sum_{k_C+k_D=k}\frac{k!}{k_{C}!k_{D}!}(q_{C|C})^{k_{C}}(q_{D|C})^{k_{D}} \cdot \frac{k_{D}g_{D}}{k_{C}g_{C}+k_{D}g_{D}}.
\end{eqnarray}
\end{small}
\noindent Then, the time evolution of $\phi_{C}$ is given by 
\begin{eqnarray}
\frac{d\phi_{C}}{dt} = \sum_{k=1}^{N-1}\frac{k}{zN}\Pr(\Delta \phi_{C}=\frac{k}{zN}) - \sum_{k=1}^{N-1}\frac{k}{zN}\Pr (\Delta \phi_{C}=-\frac{k}{zN}). 
\label{eq:27}
\end{eqnarray}

Similar to previous subsection, we can also derive the quasi-steady relations: $(q_{C|C}-q_{C|D})=\frac{1-q}{z-1}$, $\phi_{C}=\phi_{k,C}$ and $\phi_{D}=\phi_{k,D}$. Using these relations the simplified form of  Eq.(\ref{eq:27}) is given by the following expression:
\begin{eqnarray}
\frac{d\phi_{C}}{dt} & = & \frac{\omega}{N}(\frac{z+q-2}{z-1})\phi_{C}(1-\phi_{C})[z(\pi_{CD}-\pi_{DD})+(1-q)(\pi_{CC}-\pi_{CD}+\frac{\alpha}{z}) \nonumber \\
& & +\alpha(1+\frac{1}{z})(z+q-2)\phi_{C}]+\mathcal{O}(\omega^2),
\end{eqnarray}
\noindent where $\alpha=\pi_{CC}-\pi_{CD}-\pi_{DC}+\pi_{DD}$, and the influential time-factor is equal to $\frac{\omega}{N}(\frac{z+q-2}{z-1})$. On consideration of low diffusion rate, i.e., neglecting the term $\frac{q}{z}$, taking $t^{\prime}=\frac{\omega}{N}(\frac{z+q-2}{z-1})t$ and using the transition probabilities related to Moran process, the modeling formalism brings out the third-order standard replicator equation with  the transformed payoff matrix:
\begin{center}
$\bordermatrix{~ & C & D \cr
                  C & (z-\frac{1-q}{z})(\pi_{CC}-\pi_{DC}) & (z-\frac{1-q}{z})(\pi_{CD}-\pi_{DD}) \cr
                  D & (q-1-\frac{1-q}{z})(\pi_{DC}-\pi_{DD}) & (q-1-\frac{1-q}{z})(\pi_{CC}-\pi_{CD}) \cr}$.
                  
\end{center}
\noindent Same intention is to be held here also with the note that the condition of a stable equilibrium is interpreted by $\alpha<0$ while $\alpha>0$ is the unstable condition of an internal equilibrium. Under the both updating processes, Birth-Death and Death-Birth, more logical forms of the replicator dynamics equations on graphs are obtained than the known equations of that (see \citep{ohtsuki2006replicator}). Here the step by step procedure of approximation calculations is very carefully handled.

Before going into the model analysis, in order to select the procedure to determine the equilibrium strategies a question can be asked: Is there any difference between the concept of evolutionary stability in a well-mixed population and for a network selection? The simple answer is {\it{yes}}. The evolutionarily stable strategy (ESS) and the strict Nash equilibrium are both dependent on the original payoff matrix in the well-mixed population, while not only the ESS but also the strict Nash equilibrium are to be influenced by the transformed payoff matrix; basically both are to be calculated on the transformed payoff matrix which governs the Moran-evolution in structured population; and author claims that concept-wise there is no difference between ESS and strict Nash equilibrium (compare with \citep{ohtsuki2008evolutionary}). We know that a monomorphic population of a strict Nash equilibrium rejects the invasion of any other strategy; the fact justifies the term equilibrium. Throughout the paper, we consider the strict Nash equilibrium -- traditional stability concept.

\section{Comparative analysis between the theoretical predictions and the stochastic simulations}
\label{sec:4}
Now, it is the stage to specify the game environments. In the consideration of a two-player version of the interaction, we  utilise the following payoff matrix for the row player \citep{zukewich2013consolidating}

\begin{center}
$\bordermatrix{~ & C & D \cr
                  C & \frac{b}{2}(1+v)-c & \frac{b}{2}-c \cr
                  D & \frac{b}{2} & 0 \cr}$,
                  
\end{center}
\noindent where  in a common structural framework, cooperators pay cost $c>0$ to provide a benefit $b>c$. The benefit is equally split between the two players, regardless of their strategies in the influence of the two weighted factors  one and $v$. Clearly, the benefits are synergistically enhanced if $v>1$ and  if $v<1$ loss of benefits is accounted. Defectors neither pay cost nor contribute benefit. Based on this rule, in order to encompass general social dilemmas, we normalise the payoff matrix by adding $c$, then dividing by $b/2$. Under the constraints: $c>0$ and $b/2>1$, the rescaling procedure reduces selection strength; the deed is feasible because we are focusing on the weak selection limit. Hence, after the rescaling and introducing the cost-benefit ratio, $u=2c/b$, the form of normalized payoff matrix is given by: 
\begin{center}
$\bordermatrix{~ & C & D \cr
                  C & 1+v & 1 \cr
                  D & 1+u & u \cr}$.
                  
\end{center}
\noindent The characters of the four fundamental social dilemmas are explained through the following relations between the two game parameters $u$ and $v$:-  Prisoner's Dilemma: $1+v>u>1, u>v$; Stag-Hunt Game: $v>u>1$; Snowdrift Game: $1>u>v$; Byproduct Mutualism: $u<1,u<v$. We capture these four social dilemmas in the well-known three game classes, namely, Dominance game, Coexistence game and Coordination game.

\subsection{Moran-evolution in well-mixed populations}
\label{subsec:4.1}
In this subsection, the evolution of cooperation is studied through the comparison between the deterministic structure and the stochastic structure. Here, we use the two formalisms for stochastically describing the time behavior of cooperation population, one is Langevin stochastic framework structure, which is directly related to the master equation and another is a Gillespie stochastic framework structure which uses a rigorously derived Monte Carlo procedure to numerically simulate the time evolution of the given reaction system. In the section \ref{sec:1}, we have already mentioned the reaction system that addresses the Moran process appropriately. And, now, it is important to define the propensity of each reaction in order to achieve Gillespie-algorithmic version of Moran-Evolution. Here, the propensity of the reaction of creation of a cooperator is defined as: average birth strength of cooperators on the fraction of the death of defectors in the particular infinitesimal time interval, this is equal to $T_{i}^{+} \times \mbox{number of coopetators}$. Similarly, the propensity of the reaction of destruction of a cooperator is defined as: average birth strength of defectors on the fraction of the death of cooperators in the particular infinitesimal time interval, this is equal to $T_{i}^{-} \times \mbox{number of defectors}$.

It would be relevant to point out that as in the deterministic structure the Moran-evolution of cooperation population is directly proportional to the intensity of the selection, $\omega$ and as the model structure is based on the weak selection framework, the overall evolution process is quite slow -- which happens in the natural world. Moreover, the dynamics in Gillespie stochastic framework structure is slower than the deterministic structure, inherently. Reason for that we capture both the dynamics at the same time coordinate system through the relation of: $t_{\mbox{deterministic}}= \omega \times t_{\mbox{stochastic}}$. Here, the comparison is made in view of that whether the trajectories of two dynamics in the different measurement patterns follow the same path or not. However, as the deterministic equation is derived from the Langevin equation, there is no time lag between two time scales related to the Langevin and deterministic pair. Additionally, in the following analysis, it is to be noted that the nature of randomness of Langevin stochastic dynamics is relatively lower than the Gillespie stochastic dynamics which is also expected according to their mathematical formalisms.  
\sloppy
\begin{figure}[t!]
\makebox[\linewidth][c]{
\begin{minipage}{.5cm}
\rotatebox{90}{\scriptsize{Proportion of Cooperators}}
\end{minipage} 
\hspace{-0.45cm}
\begin{minipage}{.32\linewidth}
\centering
\scriptsize{Dominance Game} \\[10pt]
\includegraphics[width=0.9\linewidth]{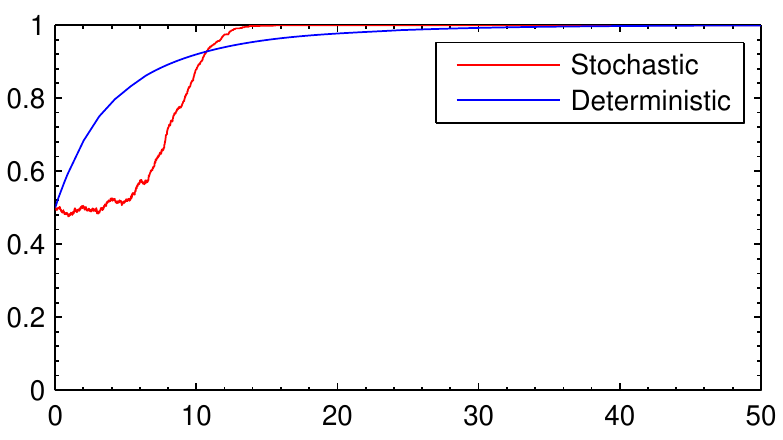}
\label{fig:1}
\end{minipage} 
\begin{minipage}{0.32\linewidth}
\centering
\scriptsize{Coexistence Game} \\[10pt]
\includegraphics[width=0.9\linewidth]{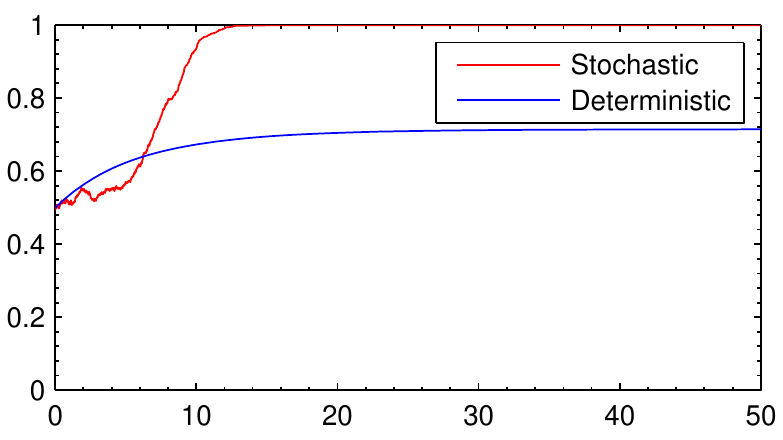}
\label{fig:2}
\end{minipage} 
\begin{minipage}{0.32\linewidth}
\centering
\scriptsize{Coordination Game} \\[10pt]
\includegraphics[width=0.9\linewidth]{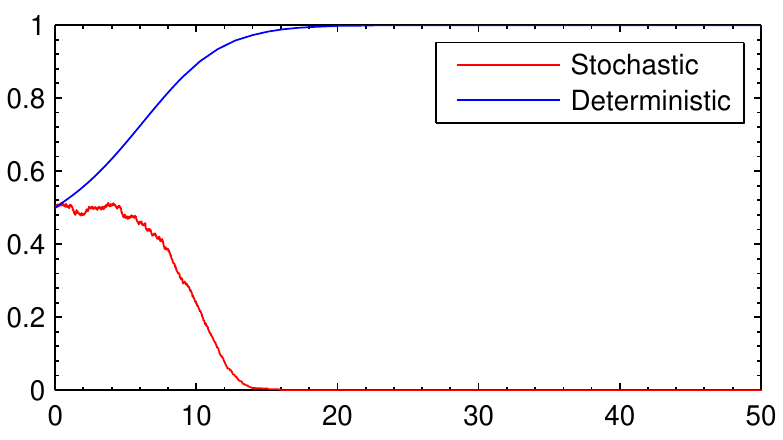}
\label{fig:3}
\end{minipage}} 
\centering
\scriptsize{Time} \\
\caption{ Moran-evolution of cooperation in the deterministic framework structure and in the Gillespie stochastic framework structure. Draw attention to the deviation between the deterministic structure and the stochastic structure. The value of population size, the initial value of the population size of cooperators, and the value of intensity of selection, among the three games, are $N=1000$, $i=500$, $\omega=0.01$, respectively. The sets of the game parameter values from left to right are:- Byproduct Mutualism game: $(u,v)=(0.2,0.3)$, Snowdrift game: $(u,v)=(0.5,0.3)$ and Stag-Hunt game: $(u,v)=(1.2,1.6)$. (For interpretation of the references to colour in this figure legend, the reader is referred to the web version of this article.)}\label{fig:GA}
\end{figure}
\begin{figure}[b!]
\makebox[\linewidth][c]{
\begin{minipage}{.5cm}
\rotatebox{90}{\scriptsize{Proportion of Cooperators}}
\end{minipage} 
\hspace{-0.45cm}
\begin{minipage}{.32\linewidth}
\centering
\scriptsize{Dominance Game} \\
\tiny{(Byproduct Mutualism)}\\[10pt]
\includegraphics[width=.9\linewidth]{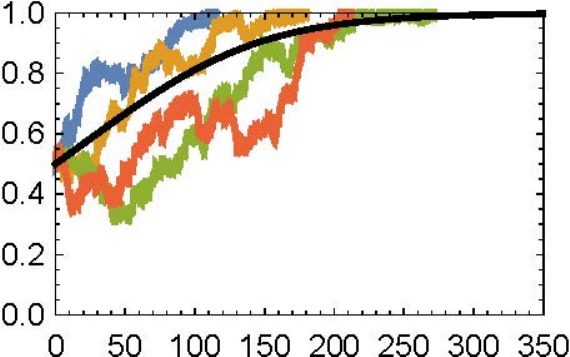}
\label{fig:4}
\end{minipage} 
\begin{minipage}{0.32\linewidth}
\centering
\scriptsize{Dominance Game} \\
\tiny{(Prisoner's Dilemma)}\\[10pt]
\includegraphics[width=.9\linewidth]{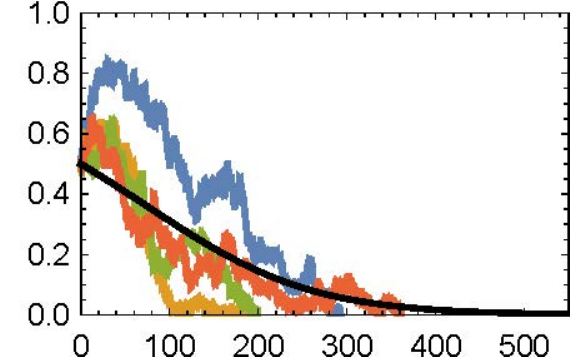}
\label{fig:5} 
\end{minipage}}\\[15pt] 
\makebox[\linewidth][c]{
\begin{minipage}{.5cm}
\rotatebox{90}{\scriptsize{Proportion of Cooperators}}
\end{minipage} 
\hspace{-0.45cm}
\begin{minipage}{0.32\linewidth}
\centering
\scriptsize{Coexistence Game}\\
\tiny{(Snowdrift Game)}\\[10pt]
\includegraphics[width=.9\linewidth]{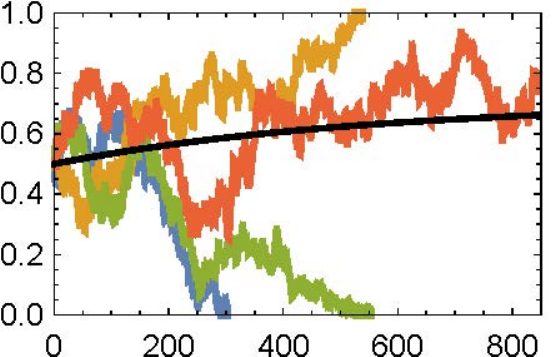}
\label{fig:6}
\end{minipage} 
\begin{minipage}{0.32\linewidth}
\centering
\scriptsize{Coordination Game} \\
\tiny{(Stag-Hunt Game)}\\[10pt]
\includegraphics[width=.9\linewidth]{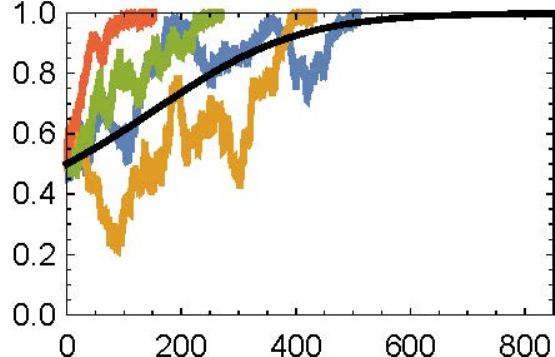}
\label{fig:7}
\end{minipage}} 
\centering
\scriptsize{Time} \\
\caption{Moran-evolution of cooperation in the deterministic framework structure and in the Langevin stochastic framework structure. Draw attention to the deviation between the deterministic structure and the stochastic structure. The stochastic trajectories are fluctuated around the black deterministic trajectory. The value of population size, the initial value of the population size of cooperators, and the value of intensity of selection, among the three games, are $N=1000$, $i=500$, $\omega=0.01$, respectively, same as Fig.\ref{fig:GA}. The sets of the game parameter values from left to right and then from top to bottom are:- $(u,v)=(0.2,2)$, $(u,v)=(2.2,2)$, $(u,v)=(0.5,0.3)$ and $(u,v)=(1.1,2)$. (For interpretation of the references to colour in this figure legend, the reader is referred to the web version of this article.)}\label{fig:LE}
\end{figure}
\sloppy

Following the three scenarios defined by the three different payoff matrices corresponding to the three game classes, the model analysis is performed. We start with the dominance game to be interpreted as either $\pi_{CC}>\pi_{DC}$ and $\pi_{CD}>\pi_{DD}$, expressing that $C$ individual is favourable than $D$ individual, or $\pi_{DC}>\pi_{CC}$ and $\pi_{DD}>\pi_{CD}$, showing the opposite fact that $D$ individual dominates $C$ individual. Both Byproduct Mutualism (BM) and Prisoner's Dilemma (PD) belong to the category of dominance game. In the Byproduct Mutualism (BM) the fitness strength of $C$ individual is invariably higher than $D$ individual, whereas in the Prisoner's Dilemma (PD) $D$ individual invariably gains higher payoff than $C$ individual. In this arena, the mentioned outcomes of the dynamical model are obtained  by varying the initial size of cooperators and the model parameters through the following ranges: $500 \leq i \leq 600$, $0.01 \leq \omega \leq 0.001$, $0<u<v<2.5$ meets $u<1$ for BM game and $1<u<1+v<3.5$ meets $u>v$ for PD game. In the most of the cases, it is assumed that at the initial state of the process, the population consists of $50$\% of individuals with the $C$ strategy and remaining $50$\% of individuals with the $D$ strategy so that either of the strategies does not get the initial advantage. With this initial value, at the  end of the process, we do not encounter any deviation from the value of $\rho_{C}=1$ for BM game and from the value of $\rho_{D}=1$ for PD game in the deterministic framework structure while, on the contrary, the deviation about the expected value $1$ of the fixation probabilities of $C$ and $D$ is observed over a specific number of different stochastic structure realizations, and the value of this deviation is higher in Gillespie stochastic structure than the Langevin stochastic structure. Moreover, in the setting range Gillespie stochastic structure is unable to show the natural character of evolutionary PD game dynamics. Whatever the values of parameters we use, we can observe the deviation of the fixation probability; however, it is seen that after the increase of certain value of the initial size of cooperators (defectors in the case of PD game) the fixation-probability-fluctuation is disappeared. The most noticeable observation comes next as the flip side of the preceding observation when the stochastic structure shows the increase in the strength of the deviation with  decreasing of the selection strength, $\omega$. The inference is that a small  perturbation about the neutral drift makes high deviation in the value $1$ of the fixation probability where at the neutral drift, the value of the fixation probability is to be obviously equal to $0$. Thus, the aspect highlights the highly unstable character of neutral drift. In this series of observations, the last finding is that with increasing of the fitness parameter $u$ in case of the defection domination game or $v$ in case of the cooperation domination game, the fixation-probability-fluctuation over the different simulations tends to the numerical value $0$ but never attains to the limiting value properly in the mentioned ranges. The Fig.\ref{fig:GA} and Fig.\ref{fig:LE} are depicted for some specific values of the game parameter pair, these are the representatives of the overall outcomes of those specific values of the $(u,v)$ pairs which form the bunch of the different simulations; however, for the certain limitations as well as avoiding the obscurity in the figure presentation we do not show  all the curves having fluctuation about one.

We attain the second scenario through the coexistence game, when $\pi_{CC}<\pi_{DC}$ and $\pi_{CD}>\pi_{DD}$ and the game determines that in the intertype competition, both types are at a less favourable when present in great quantity, and are in a favourable when present in less quantity. Clearly, this is the criterion for the existence of an internal equilibrium point where the two types coexist. The snowdrift game (SD game) is a specific example of coexistence game. The internal equilibrium point is calculated by the well known formula of replicator dynamics, given by $x^{*}=\frac{\pi_{DD}-\pi_{CD}}{\pi_{CC}-\pi_{CD}-\pi_{DC}+\pi_{DD}}$; however, the taking time to reach the point is relatively high here. And the fluctuation at $\rho_{C}=\rho_{D}=0$ is prominently visible in the stochastic structures; also, over the prolonging simulation time range Gillespie stochastic structure fails to demonstrate the coexistence phenomenon in the given parameter constraints: $1>u>v$. The deficiency is unnoticeable with higher value of the population size. The fact suggests that sometime the high stochasticity leads unrealistic result because in the latter section we show that in a finite population structure the coexistence is relatively common; which is intuitively quite acceptable.

The coordination game pattern in which $\pi_{CC}>\pi_{DC}$ and $\pi_{CD}<\pi_{DD}$, represents the third and last scenario where the advantage linearly depends on the summation value of the payoffs of the intratype and intertype competitions in such a way for $\pi_{CC}+\pi_{CD}>\pi_{DC}+\pi_{DD}$, type $C-$individual dominates type $D-$individual and on the contrary, in the case of less than type inequality, $D$ individual is favourable. For this reason, stag-hunt game (SH game), an example of coordination games, has a stable equilibrium point either at $x^{*}=1$ or $x^{*}=0$ in the game parameter range: $2.5>v>u>1$. The fluctuation at the point $\rho_{C}=1$ has been shown in the coordination-game-panel of the Fig.\ref{fig:GA}. Regarding the computer simulation, the fluctuation at the point $\rho_{D}=1$ (or at the point $\rho_{C}=1$) is also seen over many stochastic realizations with that expectation at the end of the process, $\rho_{D}$ (or $\rho_{C}$ ) will have to take either value $1$ (or value $0$) for the defection dominance game or the value $0$ (or value $1$) for the cooperation dominance game. Similar to previous, the randomness character is higher in Gillespie structure than the Langevin stochastic structure relative to the theoretical prediction of Moran-evolution of cooperation. 

\subsection{Moran-evolution in structured populations}
\label{subsec:4.2}

\begin{figure}[t!]
\centering
\includegraphics[width=0.5\textwidth]{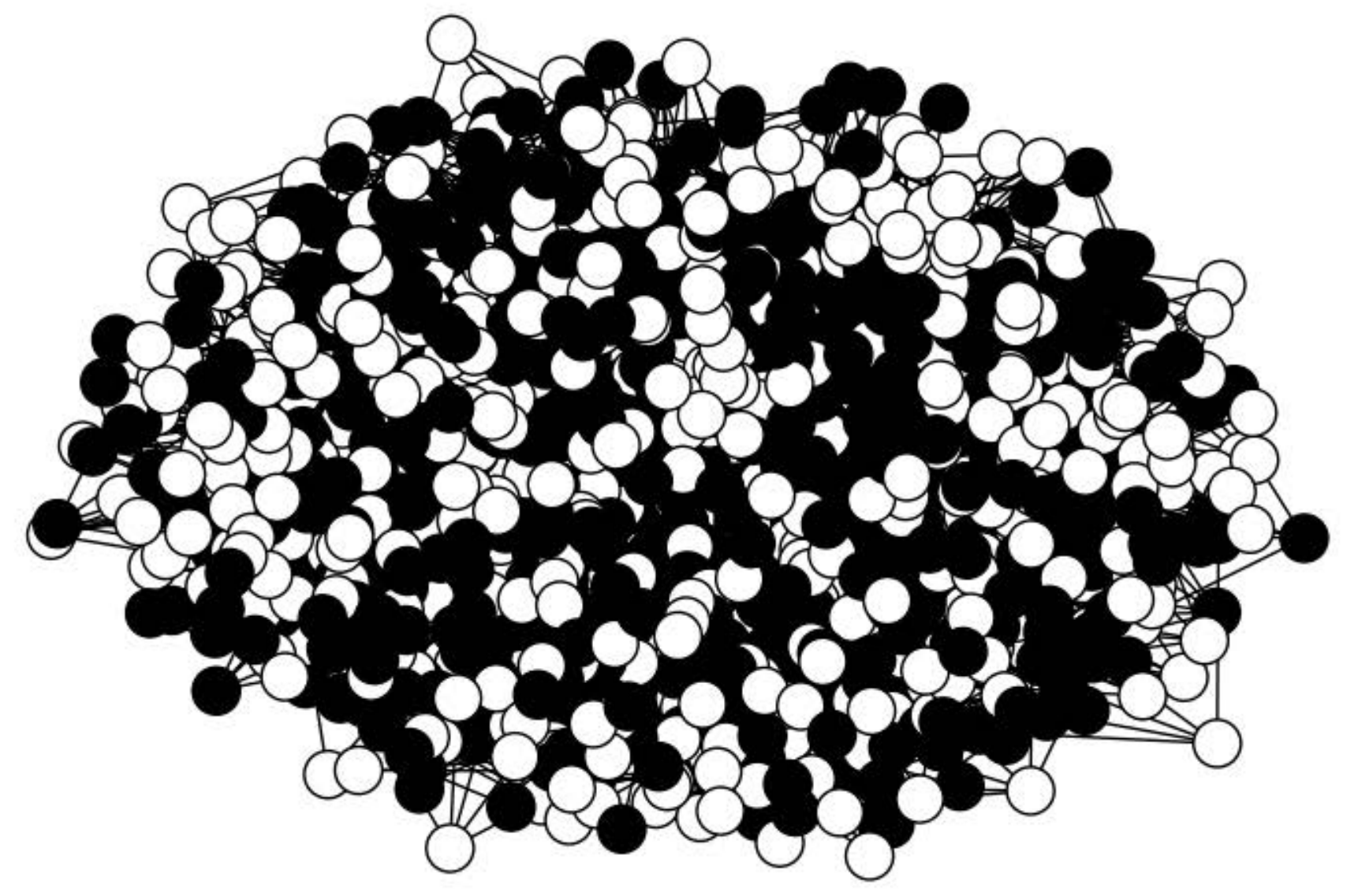}
\caption{An illustration of the one of the initial population structures represented by Barbasi-Albert network with average degree $9$. Here, in $1000$ individuals, the initial allotments of cooperators on vertices are defined through the command: random()$<$0.5, where $w=0.01$, same as previous. The general rule to determine a initial number of cooperators have been set up as random()$<$ neighbourhood of the predicted value $\phi_{C}^{*}(t,6.5)$. The solid black balls indicate the cooperators. Each update procedure do run  up to the $5\times10^4$ time steps and each simulation point of all simulations $\phi_{C}^{*}(5\times10^4,z)$ is obtained by averaging the values of $\phi_{C}(t,z)$ over the last $10,000$ time steps. Among the $10$ different network-simulation realizations, the three values of the random variable $\phi_{C}^{*}(5\times10^4,z)$ -- the most closed value along with the upper and lower ranges of deviation about the analytically predicted value -- have been plotted for each value of $z$ in a comparison framework. } \label{fig:BAz=9}
\end{figure}

We now proceed to measure the accuracy of the theoretical results by comparison with models for dynamics on networks to be characterised as the class of dynamical network models in which the network topology is fixed throughout time. Following the either of the two update rules, on the nodes-represent-individuals network, the cooperation and defection states of individuals change over time through their interaction with other individual-states those are connected to them. Here, the changes move asynchronously, like voter model. We consider games played on two commonly use families of networks: the regular graphs (RG), and the scale-free networks -- in particular the Barbasi-Albert networks (BAN). Each individual numerically takes either $0$, equating the defection or $1$, equating the cooperation (see Fig.\ref{fig:BAz=9}). Both update rules accomplish two specific steps: in one step, reproduction takes place as a consequence of an individual's fitness and in the other step, death occurs in a random way. Besides the order of these two steps, the selection of an individual depends upon the payoff matrix of the considering game and the number of connected links with the individual, and algorithmic point of view both the influential factors act on the evolution of an individual's character according to the well-known concept of preferential attachment. Reason for that it is natural to expect under either of the update rules both cooperation and defection can be beneficial or not. However, keeping in mind the analogous concept of the notable friendship paradox -- a randomly selected neighbour of a randomly selected node is likely to have a larger-than-average degree -- we can say that in the context of the cooperation dominance game, the DB update model would promote cooperation while in the BD update model, promotion of cooperation would be a slow procedure because  a chosen cooperator can replace either a cooperator or a defector that implies either the cooperators increases their number or remains unchanged in each of the iterative steps.

To start the analysis, at the beginning of this subsection, we have to point out the ranges of the feasible parameter space $(z,q,(u,v))$ of the internal equilibrium $\phi_{C}^{*}$, by using the two conditions $0\leq \phi_{C}^{*} \leq 1$ and $\alpha=v-1 \lessgtr 0$. The simple algebras provide the following forms of the stable spaces where $\alpha < 0$ :
\begin{equation*}
\begin{rcases}
  z(v-u) & \leq v-(1-q) \\
  v(q-1)+1 & \leq z(1-u) 
\end{rcases}
\text{corresponding to BD updating}
\end{equation*}
and 
\begin{equation*}
\begin{rcases}
 (1-u)+(\frac{1-q}{z}) & \leq \alpha (\frac{1}{z^2}-1) \\
  (1-q)(\alpha(1+\frac{1}{z})+1) & \geq z(u-1) 
\end{rcases}
\text{corresponding to DB updating}.
\end{equation*}
\noindent To get the unstable spaces, we have to change the directions of inequalities. Clearly, a numerical set value of $(z,q,(u,v))$ is to be feasible if it satisfies one of these four sets of inequalities.

\subsubsection{Dominance game}
\label{subsubsec:4.2.1}
Due to the limitations  of the feasible ranges, Byproduct Mutualism and Prisoner's  Dilemma are sole representative of the dominance Game in the BD update model and in the DB update model, respectively. In evolutionary BM game dynamics of well-mixed version, cooperation is a strict Nash equilibrium where the three equilibria are: $x^{*}=0$ (unstable), $x^{*}=\frac{u-1}{v-1}$ (unstable with the constraint: $0<u<v<1$), $x^{*}=1$ (stable). For the BD updating, on the flip side of the same dynamics, the cooperation is to be a strict Nash equilibrium if $v-u>\frac{q-1}{z-1}$ while on the condition $\frac{1-u}{v}<\frac{q-1}{z-1}$ the defection acts as a strict Nash equilibrium role. That is, for $v=0$ winning condition of cooperator over defector is $\frac{b}{c}>2(\frac{z-1}{1-q}$); on the contrary defector can only win over cooperator in the diffusion scenario with having a numerical value greater than one. Here, in the structured framework the internal equilibrium can play the dual characters, one is a stable character on the condition of $v<1$ and another is the one-sided unstable character on the condition of $v>1$. We use the phrase {\it{one-sided unstable character}} because on the condition of $v>1$ in the feasible region, the internal equilibrium $\phi_C^{*}$ exhibits unstable character only if $\phi_{C}(t_0)>\phi_{C}^{*}(t)$, while if $\phi_{C}(t_0)<\phi_{C}^{*}(t)$, then $\frac{d\phi_{C}(t)}{dt}>0$ implies in any circumstances $\phi_{C}(t)$ converges to $\phi_{C}^{*}(t)$. 

\begin{figure}[b!]
\makebox[\linewidth][c]{
\begin{minipage}{.5cm}
\rotatebox{90}{\scriptsize{Proportion of Cooperators}}
\end{minipage} 
\hspace{-0.45cm}
\begin{minipage}{.32\linewidth}
\centering
\scriptsize{Dominance Game} \\[10pt]
\includegraphics[width=0.9\linewidth]{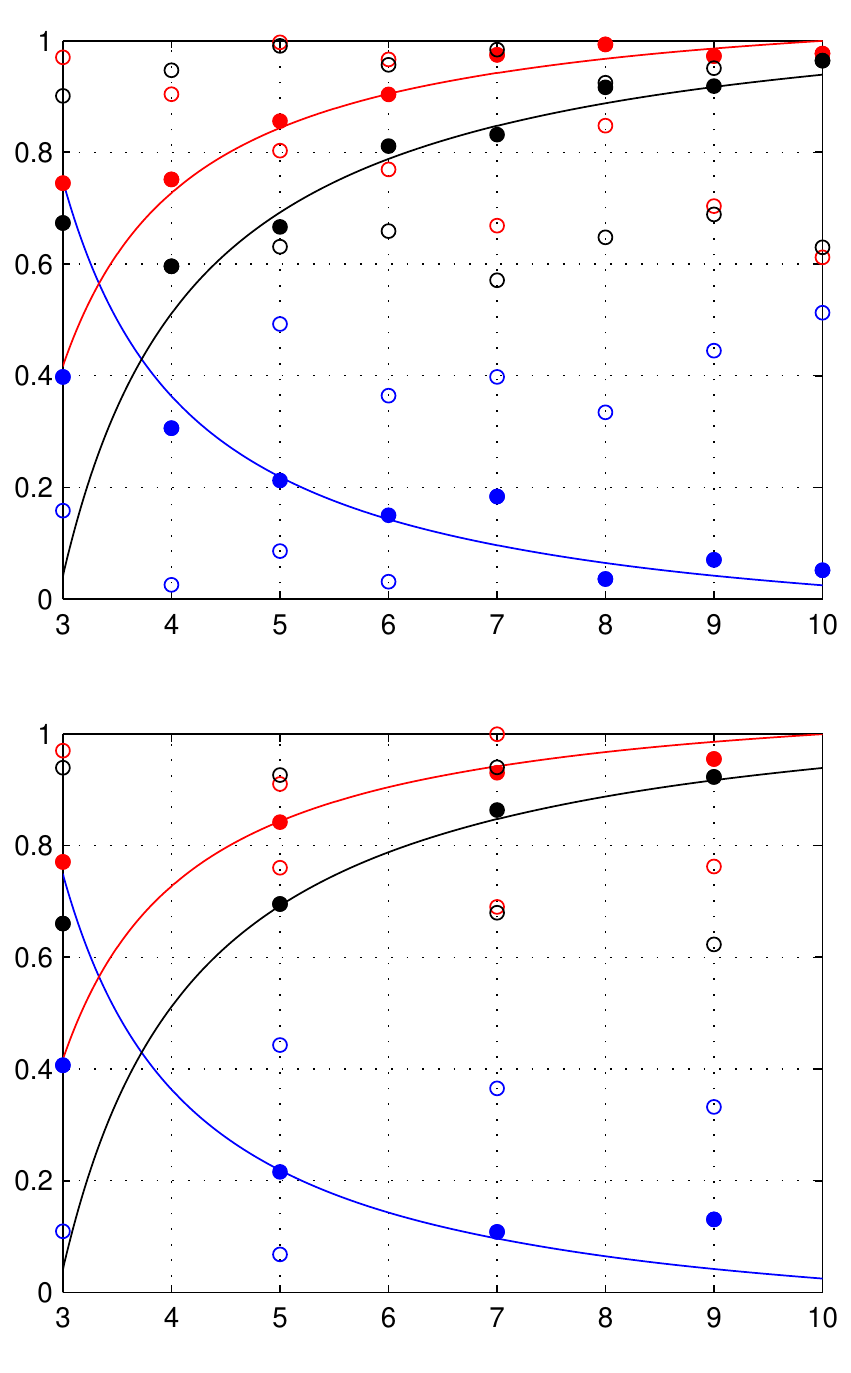}
\label{fig:41}
\end{minipage} 
\begin{minipage}{0.32\linewidth}
\centering
\scriptsize{Coexistence Game} \\[10pt]
\includegraphics[width=0.9\linewidth]{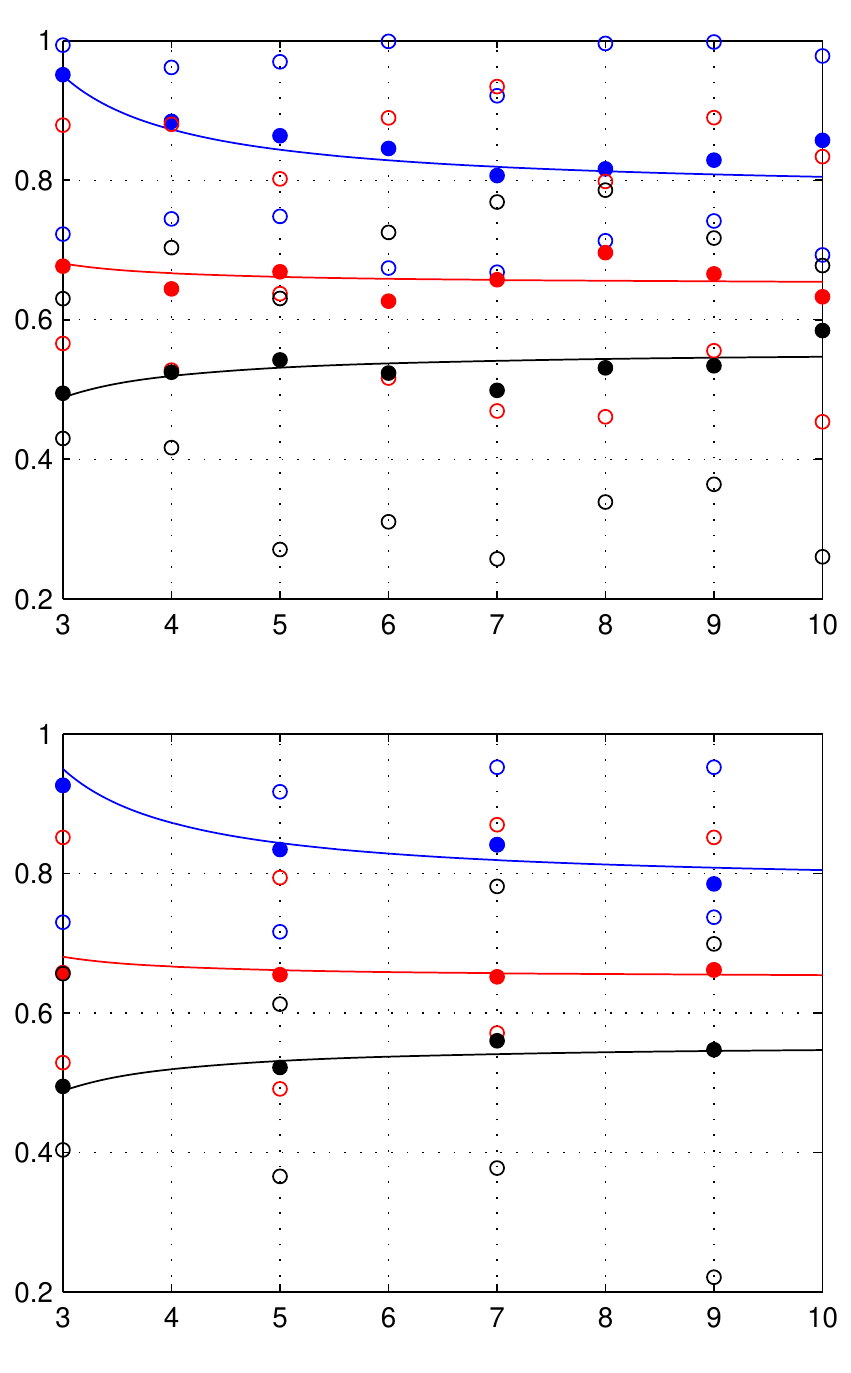}
\label{fig:42}
\end{minipage} 
\begin{minipage}{0.32\linewidth}
\centering
\scriptsize{Coordination Game} \\[10pt]
\includegraphics[width=0.9\linewidth]{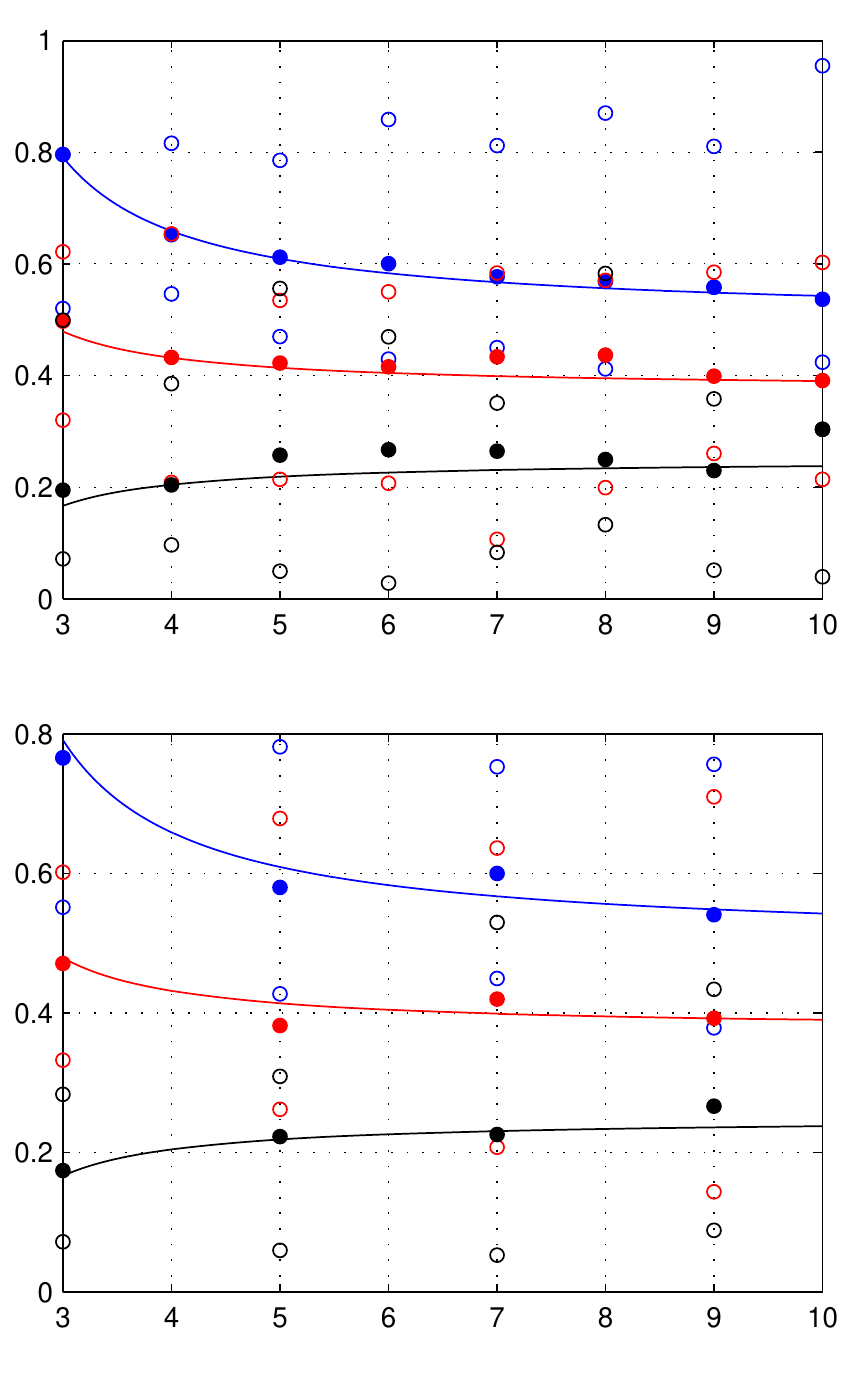}
\label{fig:43}
\end{minipage}} 
\centering
\scriptsize{(Average) Degree } \\
\caption{Moran-Evolution of cooperation through the internal equilibrium trajectory $\phi_{C}^{*}(t,z)$ under Birth-Death updating. Three panels along the first row are related to the evolution on regular graphs and other three being the evolution on Barbasi-Albert networks have been placed along the second row. In BM game dynamics, the three trajectories are assigned by the three different colours, blue: $(z, 0.2,(0.99,1.1))$ with random()$<$0.25, red: $(z, 0.2,(0.89,0.9))$ with random()$<$0.85 and black: $(z, 0.25,(0.89,0.9))$ with random()$<$0.8. Similarly, for SD game dynamics, the three trajectories correspond to the three colours those are defined as, blue: $(z, 0.2,(0.61,0.5))$ with random()$<$0.85, red: $(z, 0.2,(0.61,0.4))$ with random()$<$0.69 and black: $(z, 0.2,(0.61,0.3))$ with random()$<$0.5. And, the allotted values for the three colours in SH game dynamics are, blue: $(z, 0.2,(1.4,1.8))$ with random()$<$0.63, red: $(z, 0.2,(1.3,1.8))$ with random()$<$0.43 and black: $(z, 0.2,(1.2,1.8))$ with random()$<$0.21. (For interpretation of the references to colour in this figure legend, the reader is referred to the web version of this article.)}\label{fig:BD}
\end{figure}

To perform the quantitative analysis, we comply with the three events in all the outcomes viz.: the decreasing event in which the proportion of cooperators is decreased, the exact event in which the proportion of cooperators lies on the range $(-0.05+\phi_{C}^{*},0.05+\phi_{C}^{*})$, and the increasing event in which the proportion of cooperators is increased, where corresponding probability measurement variables  are denoted by $\mbox{P}_{d}$, $\mbox{P}_{e}$ and $\mbox{P}_{i}$, respectively, and their values have been tabulated at each value of $z$ \footnote{The calculation procedure: According to the classical definition of  probability if at $z=z_{1}$ an event can occur in $h$ different ways out of  a total number of $n$ random experiments, then the  value of the event is $\frac{h}{n}$.}; these variables characteristically are three random variables. The increase and the decrease are relative to the number of cooperators at the beginning of the Moran-evolution process. Comparison between the theoretical results and the computer simulation data that is done through figure presentation and tabulated data analysis leads to the same conclusion where cooperation is a strict Nash equilibrium (see Fig.\ref{fig:BD} and Table \ref{Tab:BM}). The deviation in the measurement of accuracy of the theoretical results is slightly large at the $z=3$ on both families of figures, whereas the deviation-value is being small at $z=6$.


\begin{table}[ht]
\centering
\caption{Evolutionary BM game dynamics at equilibrium} 
\label{Tab:BM}
\begin{adjustbox}{width=1\textwidth}
\small
\begin{tabular}{rcccccccc}
  \hline
$(\mbox{P}_d, \mbox{P}_e, \mbox{P}_i )$ & at $z=3$ &at $z=4$  & at $z=5$ & at $z=6$ & at $z=7$ & at $z=8$ &at $z=9$ & at $z=10$\\ 
  \hline
\rowcolor[gray]{0.8}\mbox{blue: RG}&(0.8, 0.0, 0.2)&(0.5, 0.1, 0.5)&(0.7, 0.5, 0.3)&(0.8, 0.6, 0.2)&(0.2, 0.0, 0.8)&(0.5, 0.2, 0.5)&(0.8, 0.1, 0.2)&(0.5, 0.5, 0.5) \\
\mbox{blue: BAN}&(0.4, 0.0, 0.6)&--&(0.6, 0.5, 0.4)&--&(0.4, 0.2, 0.6)&--&(0.8, 0.0, 0.2)&-- \\
\rowcolor[gray]{0.8}\mbox{red: RG}&(0.6, 0.0, 0.4)&(0.8, 0.4, 0.2)&(0.3, 0.5, 0.7)&(0.3, 0.5, 0.7)&(0.5, 0.2, 0.5)&(0.1, 0.3, 0.9)&(0.6, 0.2, 0.4)&(0.3, 0.1, 0.7) \\
\mbox{red: BAN}&(0.6, 0.0, 0.4)&--&(0.6, 0.3, 0.4)&--&(0.4, 0.2, 0.6)&--&(0.4, 0.5, 0.6)&-- \\
\rowcolor[gray]{0.8}\mbox{black: RG}&(0.4, 0.0, 0.6)&(0.4, 0.0, 0.6)&(0.4, 0.2, 0.6)&(0.6, 0.4, 0.4)&(0.4, 0.2, 0.5)&(0.4, 0.2, 0.6)&(0.5, 0.4, 0.5)&(0.2, 0.3, 0.8) \\
\mbox{black: BAN}&(0.3, 0.0, 0.7)&--&(0.3, 0.1, 0.7)&--&(0.2, 0.3, 0.8)&--&(0.5, 0.1, 0.5)&-- \\
   \hline
\end{tabular}
\end{adjustbox}

\end{table} 

We have seen that in the context of evolutionary PD game dynamics in well-mixed population, $C$ individual is dominated by $D$ individual. The dynamics having defection Nash equilibrium, determines two equilibria: $x^{*}=0$ (stable), $x^{*}=1$ (unstable) where the internal equilibrium $x^{*}=\frac{u-1}{v-1}$ is undefined. On the contrary, for DB updating in the feasible region of a structured population, the PD game dynamics can define the one-sided unstable internal equilibrium on the condition of $v>1$. The important note is that in the structured framework, criterion for $C$ and $D$ to be a strict Nash equilibrium respectively is : $u-v<\frac{1}{z-1}$ i.e., for $v=0$, $\frac{b}{c}>2(z-1)$ instead of $\frac{b}{c}> \mbox{(average) degree of the graph}$ and $\frac{1}{z-1}<\frac{u-1}{v}$. That is, both $C$ and $D$ can be Nash equilibria, but if at a low number of links in which $D$ exists as a Nash equilibrium strategy, then we are unable to define the internal equilibrium similar to well-mixed population structure. Basically, the possibility of coexistence of cooperators and defectors is high here with the reality that a particular strategy can dominate or can be dominated by other strategy.

\begin{figure}[t!]
\makebox[\linewidth][c]{
\begin{minipage}{.5cm}
\rotatebox{90}{\scriptsize{Proportion of Cooperators}}
\end{minipage} 
\hspace{-0.45cm}
\begin{minipage}{.32\linewidth}
\centering
\scriptsize{Dominance Game} \\[10pt]
\includegraphics[width=0.9\linewidth]{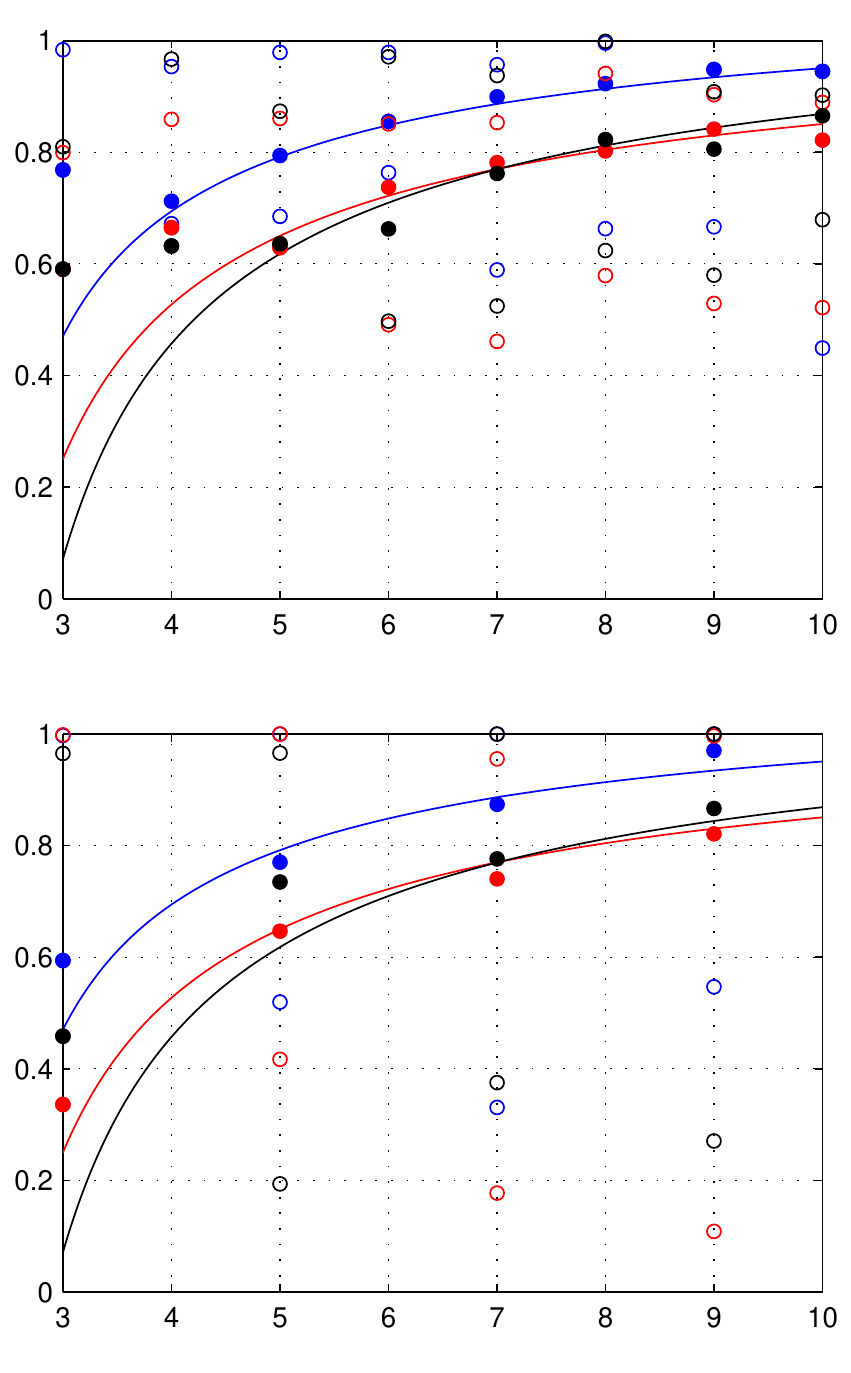}
\label{fig:51}
\end{minipage} 
\begin{minipage}{0.32\linewidth}
\centering
\scriptsize{Coexistence Game} \\[10pt]
\includegraphics[width=0.9\linewidth]{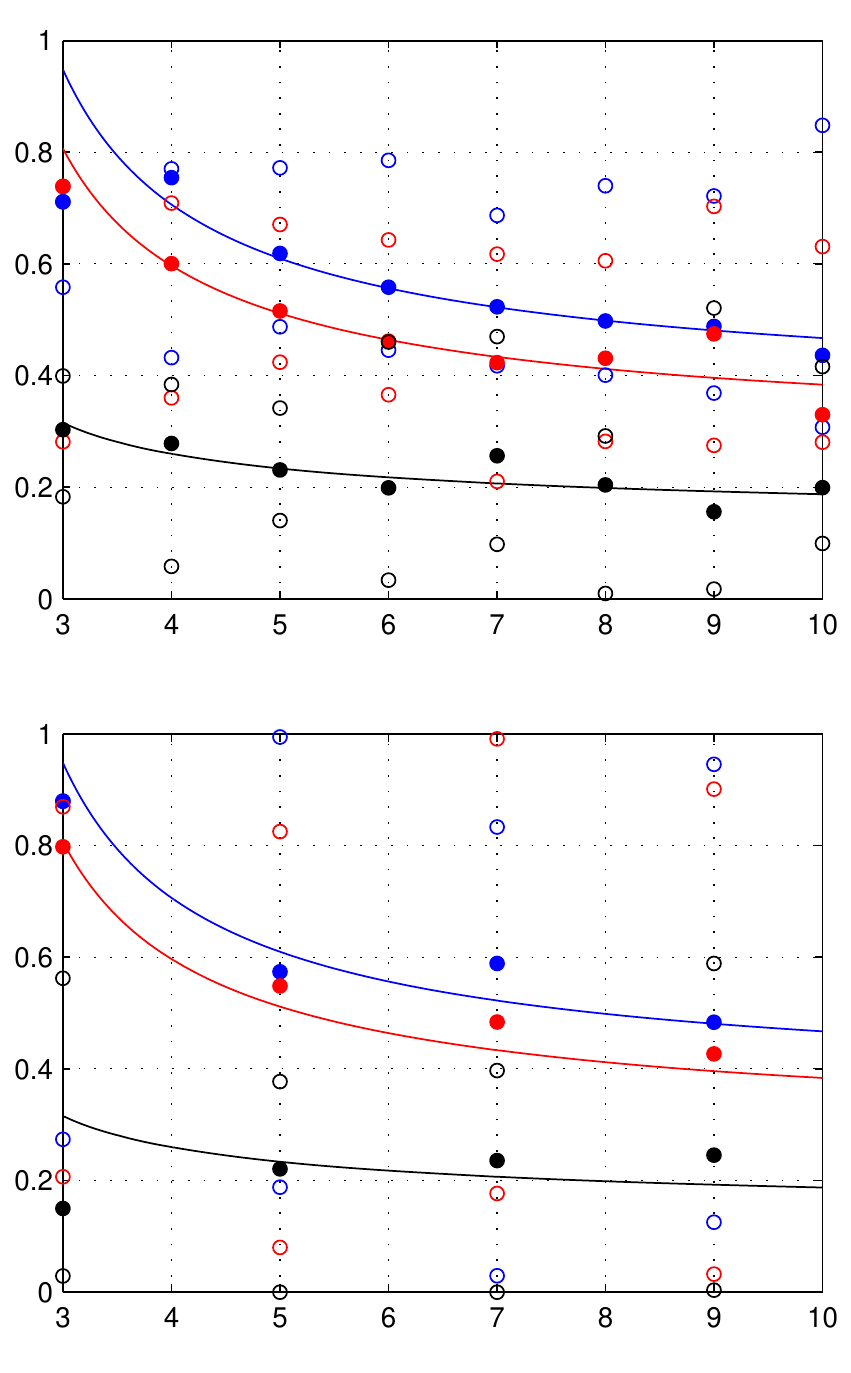}
\label{fig:52}
\end{minipage} 
\begin{minipage}{0.32\linewidth}
\centering
\scriptsize{Coordination Game} \\[10pt]
\includegraphics[width=0.9\linewidth]{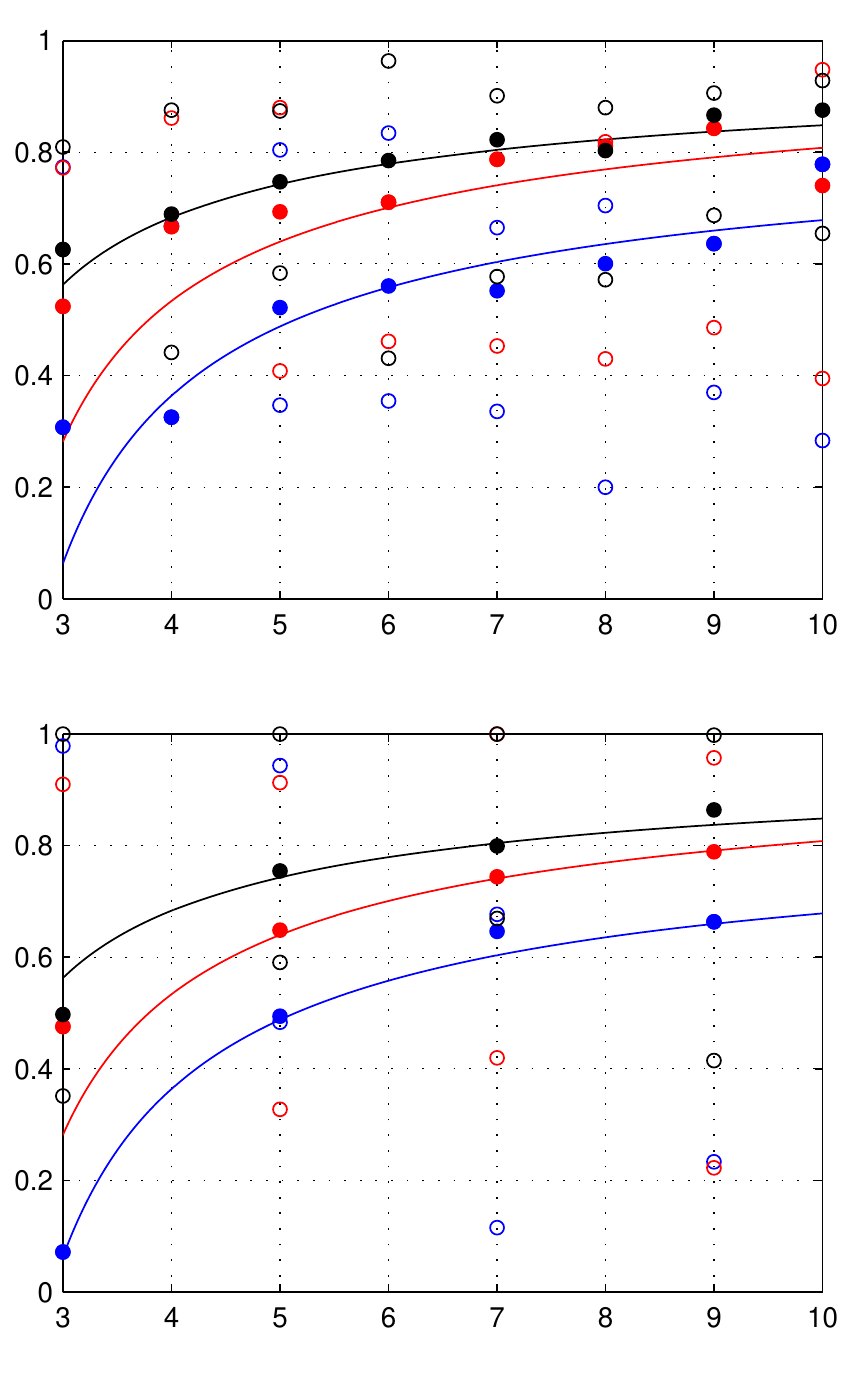}
\label{fig:53}
\end{minipage}} 
\centering
\scriptsize{(Average) Degree } \\
\caption{Moran-Evolution of cooperation through the internal equilibrium trajectory $\phi_{C}^{*}(t,z)$ under Death-Birth updating. The same panel configuration pattern of Fig.\ref{fig:BD} is followed here. In PD game dynamics, the three different colours are read as, blue: $(z,0.2,(1.6,1.55))$ with random()$<$0.83, red: $(z,0.3,(1.42,1.41))$ with random()$<$0.73 and black: $(z,0.2,(1.42,1.39))$ with random()$<$0.73. Similarly, for SD game dynamics the three colours are defined as, blue: $(z,0.2,(0.81,0.47))$ with random()$<$0.58, red: $(z,0.2,(0.85,0.47))$ with random()$<$0.5 and black: $(z,0.2,(0.9,0.3))$ with random()$<$0.23. And, the allotted values for the three colours in SH game dynamics are, blue: $(z,0.2,(1.5,1.6))$ with random()$<$0.5, red: $(z,0.2,(1.57,1.6))$ with random()$<$0.65 and black: $(z,0.4,(1.57,1.6))$ with random()$<$0.75. (For interpretation of the references to colour in this figure legend, the reader is referred to the web version of this article.)}\label{fig:DB}
\end{figure}

Same as previous, the incoherence data of $(\mbox{P}_d, \mbox{P}_e, \mbox{P}_i )$ reveal simply the character of the stochastic dynamics (see Table \ref{Tab:PD}) where on the chosen values of $(u,v)$, the strategy $C$ and strategy $D$ both are strict Nash equilibria --   except at $z=3$ and $z=4$ -- in the sense that the cooperators try to raise the rate of birth while the defectors try to survive. The result of that, whole population tends to a situation of coexistence with the high proportion value of cooperators than the defectors. Here, all the coordinate values of $(\mbox{P}_d,  \mbox{P}_i )$ also unfold this character of the evolutionary dynamics. At the beginning, $z=3$, high average numerical value of $\mbox{P}_i$ than the $\mbox{P}_d$ is the indication of cooperators' enhancement. However, this enhancement is not hundred percentages ensured. The first column in Fig.\ref{fig:DB} shows the variation of $\phi_{C}^{*}(z)$ under the Moran-evolutionary game dynamics of PD version.
\begin{table}[ht]
\centering
\caption{Evolutionary PD game dynamics at equilibrium} 
\label{Tab:PD}
\begin{adjustbox}{width=1\textwidth}
\small
\begin{tabular}{rcccccccc}
  \hline
$(\mbox{P}_d, \mbox{P}_e, \mbox{P}_i )$ & at $z=3$ &at $z=4$  & at $z=5$ & at $z=6$ & at $z=7$ & at $z=8$ &at $z=9$ & at $z=10$\\ 
  \hline
\mbox{blue: RG}&(0.6, 0.0, 0.4)&(0.5, 0.3, 0.5)&(0.6, 0.4, 0.4)&(0.3, 0.5, 0.7)&(0.4, 0.2, 0.6)&(0.4, 0.1, 0.6)&(0.7, 0.1, 0.3)&(0.3, 0.2, 0.7) \\
\rowcolor[gray]{0.8}\mbox{blue: BAN}&(0.3, 0.0, 0.7)&--&(0.5, 0.2, 0.5)&--&(0.5, 0.1, 0.5)&--&(0.5, 0.3, 0.5)&-- \\
\mbox{red: RG}&(0.4, 0.0, 0.6)&(0.2, 0.0, 0.8)&(0.5, 0.1, 0.5)&(0.5, 0.1, 0.5)&(0.4, 0.5, 0.6)&(0.6, 0.3, 0.4)&(0.3, 0.3, 0.7)&(0.5, 0.2, 0.5) \\
\rowcolor[gray]{0.8}\mbox{red: BAN}&(0.2, 0.1, 0.8)&--&(0.3, 0.1, 0.7)&--&(0.8, 0.1, 0.2)&--&(0.4, 0.1, 0.6)&-- \\
\mbox{black: RG}& (0.4, 0.0, 0.6)&(0.4, 0.0, 0.6)&(0.6, 0.3, 0.4)&(0.3, 0.1, 0.7)&(0.4, 0.4, 0.6)&(0.4, 0.3, 0.5)&(0.3, 0.1, 0.7)&(0.1, 0.3, 0.9) \\
\rowcolor[gray]{0.8}\mbox{black: BAN}&(0.3, 0.0, 0.7)&--&(0.2, 0.0, 0.8)&--&(0.4, 0.1, 0.6)&--&(0.4, 0.2, 0.6)&-- \\
   \hline
\end{tabular}
\end{adjustbox}

\end{table} 
\sloppy
\sloppy

\subsubsection{Coexistence game}
\label{subsubsec:4.2.2}
The evolutionary snowdrift game dynamics in well-mixed population -- a specific example of coexistence game dynamics -- is characterised as nether $C$ nor $D$ is a Nash equilibrium with three equilibria: $x^{*}=0$ (unstable), $x^{*}=\frac{1-u}{1-v}$ (stable), $x^{*}=1$ (unstable). For both updating, the corresponding evolutionary dynamics on structured pattern determines a stable internal equilibrium $\phi_{C}^{*}(t)$ where the natures of other two  equilibria remain also unchanged. In addition, SD dynamics retains the identical Nash equilibrium criterion. However, in case of BD updating either strategy, not both simultaneously, is a Nash equilibrium while DB updating reveals the fact that  $C$ can only be a Nash equilibrium. The sets of inequalities:  $u-v>\frac{1-q}{z-1}$, $\frac{q-1}{z-1}<\frac{1-u}{v} \}$ corresponding to BD updating and $u-v>\frac{1}{z-1}$, $\frac{1}{z-1}>\frac{u-1}{v} \}$ corresponding to DB updating, are satisfied for a wide range of parameter space $(z,q,(u,v))$, which  asserts that possibility of coexistence of cooperators and defectors is high enough. 

Here, neither $C$ nor $D$ is favourable in $z$ past the certain numerical value, that being clarified in Figs.\ref{fig:BD}-\ref{fig:DB} along the column of coexistence game. All the results of computer simulation are summarized in Table \ref{Tab:SD} in the form of $(\mbox{P}_d, \mbox{P}_e, \mbox{P}_i )$. We notice that the theoretical prediction in coexistence game is in  relatively good agreement with numerical iteration simulation than the other two game dynamics. From this observation, one can draw an important conclusion: nature always tries to create an environment where all the different characters can survive. This explanation is more supportive when we examine the coordinate values of $(\mbox{P}_d, \mbox{P}_i )$ in all three evolutionary game dynamics. Mathematically, to this place, a significant consequence is that inherently the iterative method, partly obeys the stability condition of the theoretical dynamical equation.

\begin{table}[t]
\centering
\caption{Evolutionary SD game dynamics at equilibrium} 
\label{Tab:SD}
\begin{adjustbox}{width=1\textwidth}
\small
\begin{tabular}{rcccccccc}
  \hline
$(\mbox{P}_d, \mbox{P}_e, \mbox{P}_i )$ & at $z=3$ &at $z=4$  & at $z=5$ & at $z=6$ & at $z=7$ & at $z=8$ &at $z=9$ & at $z=10$\\ 
  \hline
\rowcolor[gray]{0.8}\mbox{blue: RG}&(0.3, 0.5, 0.7)&(0.4, 0.3, 0.6)&(0.2, 0.6, 0.8)&(0.4, 0.3, 0.6)&(0.7, 0.4, 0.3)&(0.4, 0.2, 0.6)&(0.4, 0.3, 0.6)&(0.1, 0.0, 0.9) \\
\mbox{blue: BAN}&(0.6, 0.2, 0.4)&--&(0.6, 0.3, 0.4)&--&(0.1, 0.5, 0.9)&--&(0.2, 0.1, 0.8)&-- \\
\rowcolor[gray]{0.8}\mbox{red: RG}&(0.4, 0.2, 0.6)&(0.3, 0.1, 0.7)&(0.3, 0.3, 0.7)&(0.3, 0.1, 0.7)&(0.4, 0.3, 0.6)&(0.4, 0.1, 0.6)&(0.5, 0.2, 0.5)&(0.4, 0.1, 0.6) \\
\mbox{red: BAN}&(0.3, 0.1, 0.7)&--&(0.7, 0.3, 0.3)&--&(0.4, 0.4, 0.6)&--&(0.3, 0.4, 0.7)&-- \\
\rowcolor[gray]{0.8}\mbox{black: RG}&(0.3, 0.3, 0.7)&(0.4, 0.2, 0.6)&(0.4, 0.3, 0.6)&(0.3, 0.1, 0.7)&(0.6, 0.2, 0.4)&(0.3, 0.2, 0.7)&(0.5, 0.2, 0.5)&(0.5, 0.1, 0.5)\\
\mbox{black: BAN}&(0.7, 0.6, 0.3)&--&(0.6, 0.4, 0.4)&--&(0.6, 0.1, 0.4)&--&(0.7, 0.2, 0.3)&-- \\
\hline 
\mbox{blue: RG}&(0.2, 0.0, 0.8)&(0.7, 0.1, 0.3)&(0.4, 0.3, 0.6)&(0.3, 0.1, 0.7)&(0.4, 0.1, 0.6)&(0.5, 0.3, 0.4)&(0.4, 0.3, 0.6)&(0.6, 0.1, 0.4) \\
\rowcolor[gray]{0.8}\mbox{blue: BAN}&(0.5, 0.0, 0.5)&--&(0.7, 0.1, 0.3)&--&(0.5, 0.0, 0.5)&--&(0.4, 0.2, 0.6)&-- \\
\mbox{red: RG}&(0.7, 0.0, 0.3)&(0.2, 0.3, 0.7)&(0.2, 0.4, 0.8)&(0.6, 0.3, 0.4)&(0.5, 0.2, 0.5)&(0.8, 0.2, 0.2)&(0.5, 0.0, 0.5)&(0.4, 0.0, 0.6) \\
\rowcolor[gray]{0.8}\mbox{red: BAN}&(0.2, 0.3, 0.8)&--&(0.6, 0.1, 0.4)&--&(0.6, 0.0, 0.4)&--&(0.5, 0.2, 0.5)&--  \\
\mbox{black: RG}&(0.4, 0.3, 0.6)&(0.6, 0.2, 0.4)&(0.6, 0.3, 0.4)&(0.7, 0.4, 0.3)&(0.2, 0.0, 0.8)&(0.8, 0.2, 0.2)&(0.5, 0.1, 0.5)&(0.9, 0.5, 0.1) \\
\rowcolor[gray]{0.8}\mbox{black: BAN}&(0.9, 0.0, 0.1)&--&(0.8, 0.3, 0.2)&--&(0.5, 0.2, 0.5)&--&(0.3, 0.1, 0.7)&--  \\
\hline
\end{tabular}
\end{adjustbox}

\end{table} 

\subsubsection{Coordination game}
\label{subsubsec:4.2.3}
In the category of coordination game dynamics in well-mixed population, the Stag-Hunt game dynamics having $C$ and $D$ strict Nash equilibria, yields only one unstable  equilibrium at the internal position. Similar to previous, this equilibrium becomes a one-sided unstable equilibrium point $\phi_{C}^{*}(t)$ of structural SH dynamics. We also note that in case of both updating rules, both strategies can act as strict Nash equilibria on the wide parameter range. As the bistable character of SH dynamics is unnoticed on the structured patten, here the concept of risk dominant as well as pareto-efficient is inapplicable. We can say that the basin of attraction of strategy $C$ depends on the increasing and decreasing nature of the internal equilibrium trajectory $\phi_{C}^{*}(t,z)$. On the contrary, if there is no internal equilibrium, then only we can define a basin of attraction for strategy $D$. This is the common phenomenon for all the three game dynamics on the structured pattern.

Since both characters are strict Nash equilibria, under BD updating, after a defector birth taking place which is proportional to his/her parent's fitness the defector will have enough opportunity to replace a cooperator  at high values of $z$  and this possibility increases with increasing of $z$, which is the explanation of decreasing character of trajectory $\phi^{*}_{C}(t,z)$, that being shown in Fig.\ref{fig:BD} along the third column. To see the other trajectories, a conclusion can draw -- which is a well-know result -- that BD updating in general would not favour cooperation in the situation where both characters can flourish. On the contrary, in the same situation DB updating promotes the cooperation; the outcome is exhibited along the third column in Fig.\ref{fig:DB}. After a defector death taking place, a cooperator having high fitness value will have enough opportunity to fill up the vacant place by her offspring and this possibility increases with increasing of $z$. However, the major factor of character's promotion is the set value of $(z,p,(u,v))$ which is self-explanatory by the values of Table \ref{Tab:SH}, where  under DB updating, the average value of $\mbox{P}_{i}$ is greater than $\mbox{P}_{d}$ upto $z=7$ while under BD updating, the average difference of $\mbox{P}_{i}$ and $\mbox{P}_{d}$ is small over all values of $z$, that supports the decreasing fact of a trajectory in appropriate cases.  

\begin{table}[ht]
\centering
\caption{Evolutionary SH game dynamics at equilibrium} 
\label{Tab:SH}
\begin{adjustbox}{width=1\textwidth}
\small
\begin{tabular}{rcccccccc}
  \hline
$(\mbox{P}_d, \mbox{P}_e, \mbox{P}_i )$ & at $z=3$ &at $z=4$  & at $z=5$ & at $z=6$ & at $z=7$ & at $z=8$ &at $z=9$ & at $z=10$\\ 
  \hline
\rowcolor[gray]{0.8}\mbox{blue: RG}&(0.4, 0.1, 0.6)&(0.3, 0.4, 0.7)&(0.7, 0.2, 0.3)&(0.5, 0.2, 0.5)&(0.6, 0.3, 0.4)&(0.3, 0.2, 0.7)&(0.3, 0.1, 0.7)&(0.3, 0.2, 0.7) \\
\mbox{blue: BAN}&(0.3, 0.2, 0.7)&--&(0.6, 0.2, 0.4)&--&(0.4, 0.2, 0.6)&--&(0.7, 0.3, 0.3)&-- \\
\rowcolor[gray]{0.8}\mbox{red: RG}&(0.3, 0.3, 0.7)&(0.6, 0.1, 0.4)&(0.5, 0.3, 0.5)&(0.4, 0.3, 0.6)&(0.5, 0.1, 0.5)&(0.5, 0.1, 0.5)&(0.5, 0.2, 0.5)&(0.6, 0.4, 0.4) \\
\mbox{red: BAN}&(0.6, 0.1, 0.4)&--&(0.4, 0.1, 0.6)&--&(0.4, 0.3, 0.7)&--&(0.5, 0.3, 0.5)&-- \\
\rowcolor[gray]{0.8}\mbox{black: RG}&(0.3, 0.3, 0.5)&(0.3, 0.3, 0.7)&(0.4, 0.1, 0.6)&(0.6, 0.1, 0.4)&(0.6, 0.2, 0.4)&(0.5, 0.1, 0.5)&(0.7, 0.3, 0.3)&(0.9, 0.0, 0.1)\\
\mbox{black: BAN}&(0.6, 0.2, 0.4)&--&(0.4, 0.6, 0.6)&--&(0.3, 0.2, 0.7)&--&(0.5, 0.2, 0.5)&-- \\
\hline
\mbox{blue: RG}&(0.4, 0.0, 0.6)&(0.5, 0.1, 0.5)&(0.5, 0.2, 0.5)&(0.3, 0.4, 0.7)&(0.7, 0.0, 0.3)&(0.7, 0.1, 0.3)&(0.5, 0.2, 0.5)&(0.6, 0.0, 0.4) \\
\rowcolor[gray]{0.8}\mbox{blue: BAN}&(0.5, 0.2, 0.5)&--&(0.1, 0.2, 0.9)&--&(0.7, 0.1, 0.3)&--&(0.6, 0.1, 0.4)&-- \\
\mbox{red: RG}&(0.5, 0.0, 0.5)&(0.0, 0.0, 1.0)&(0.3, 0.1, 0.7)&(0.5, 0.2, 0.5)&(0.7, 0.1, 0.3)&(0.5, 0.0, 0.5)&(0.4, 0.0, 0.6)&(0.5, 0.0, 0.5) \\
\rowcolor[gray]{0.8}\mbox{red: BAN}&(0.3, 0.0, 0.7)&--&(0.3, 0.1, 0.7)&--&(0.2, 0.3, 0.8)&--&(0.4, 0.2, 0.6)&--  \\
\mbox{black: RG}&(0.4, 0.0, 0.6)&(0.6, 0.2, 0.4)&(0.4, 0.1, 0.6)&(0.6, 0.2, 0.4)&(0.2, 0.4, 0.8)&(0.3, 0.1, 0.7)&(0.5, 0.1, 0.5)&(0.7, 0.2, 0.3) \\
\rowcolor[gray]{0.8}\mbox{black: BAN}&(0.4, 0.0, 0.6)&--&(0.1, 0.3, 0.8)&--&(0.1, 0.3, 0.9)&--&(0.5, 0.1, 0.5)&--  \\
\hline
\end{tabular}
\end{adjustbox}

\end{table} 

\section{Discussion}
Considering the update rules, the evolution of cooperation is portrayed through the medium of the replicator equations of cooperation in  well-mixed and structured populations, in respect of the results compared with the stochastic simulations. Between the two procedures, analytic and stochastic, it is very hard to tell which one is more appropriate to draw the dynamics than the other. One can argue that stochasticity put realism in models because any natural dynamics never run properly along the predictable path and hence stochasticity should be considered to make the model realistic. However, as the modelling framework offered by computer simulation in general is limited to individual-based  stochastic simulations, the generated results naturally can deviate to true generality and the diverging outcomes of the time consuming procedure often raises its questionable validity. The fact is that real scenario can only be captured in the comparison mode in a qualitative as well as  quantitative agreement which is demonstrated throughout this research article. The detailed analysis not only helps us to present some new results but also  provides us many evidences to define the disparities -- about the cooperation enhancement condition -- among the various games on structured populations under the setting of Birth-Death and Death-Birth updating rules, those are reported in earlier literatures (see \citep{konno2011condition, iyer2016evolution}, and references therein).

We know that the dynamics represented by the replicator equation on graphs is the combined effects of two states: the equilibrium state in the dynamics of local frequencies of an $i^{\mbox{th}}$ strategist next to a $j^{\mbox{th}}$ strategist, $q_{i|j}$, and the dynamics of global frequency $\phi_{C}$, depending on the $q_{i|j}$s. To get rid the dependency barrier, it is assumed that the global frequency is in the unchanged state while the local frequencies equilibrate. This assumption is the backbone of the dynamics on graphs. Ohtsuki and Nowak \cite{ohtsuki2006replicator} calculated the equilibrium local frequencies as: $q_{C|C}^{*}=\frac{k-2}{k-1}x_{C}+\frac{1}{k-1}, q_{C|D}^{*}=\frac{k-2}{k-1}x_{C} \} $, where global frequency $x_{C}$ is constant in the quasi-steady-state. In such construction, $x_{C}$ plays dual characters over time, at one time, in the expression of $q_{i|j}^{*}$ it is constant and on the other times, after substituting it in the place of $q_{i|j}^{*}$ in the main equation of the global frequency, it is taken as a variable. The construction can be acceptable in certain point of view. However, to avoid this ambiguity, in the present article, the equilibrium local frequencies are considered as: $q_{C|C}^{*}= something+\frac{1}{z-1}, q_{C|D}^{*}= something \} $, for $q=0$; and the term $something$ is eliminated through the using of the following relation: $q_{C|C}^{*}-q_{C|D}^{*}= \frac{1}{z-1} $, which is the well-known  quasi-steady-state in the structured framework. We calculate the condition following the traditional way. Here, the robustness of calculation structure is measured by noting that $x^{*}=\lim_{z\rightarrow N-1}\phi_{C}^{*}$, for both update rules, that being expected. However, the model clearly reveals that the structured dynamics is not a linear transformation of the well-mixed dynamics.

In the consideration of evolutionary dynamics on heterogeneous network, it is great challenged to find out the correct way of modelling the updating of a population. With this intention, in order to tackle the heterogeneity, the concept of coalescent theory has been utilised in the recent research work of Allen et al. \cite{allen2017evolutionary}. The theory is implemented to calculate the benefit-to-cost ratio as a function of coalescence times of random walks. A coalescing random walk (CRW) -- a collection of random walks -- is considered on a graph G where walks move independently until two walks coalesce and the coalescence time is the expected meeting time of random walks from two distinct vertices, representing time to a common ancestor. Calculating the benefit-to-cost ratios for all possible diverse population structures, the model pattern rigidifies the known realization that cooperation flourishes most in the presence of the strong pairwise ties. Introducing the average-number-of-interactions factor Maciejewski et al.  \cite{maciejewski2014evolutionary} explained the exact same reason for the cooperation enhancement; however, their method was completely different where taken microscopic processes viz.: averaging and accumulating payoffs on the level of individuals as  foundation parts, the macroscopic features of the evolutionary process on the level of population such as frequency and distribution of cooperators were determined on heterogeneous networks. As the intention to considering the heterogeneous model is to accept the challenge to handle and to measure the influence of heterogeneity, it is natural, the question then arises as to whether the current model can read the consequence of graph heterogeneity? Before answering our question, we have to clarify a specific character of the structural pattern of Barbasi-Albert scale-free network. The growth of the model is directly dependent on the preferential attachment mechanism that being used to calculate probability $p_{i}$ which defines that the new node is connected to node $i$ with degree $k_{i}$, is: $p_{i}=\frac{k_{i}}{\sum_{j}k_{j}}$. And, it is not too hard to show through the rigorous mathematical derivation that the degree distribution $p(k)$ of scale-free network follows the power-law rule which is given by: $p(k)\sim k^{-\gamma}$, typically with $2 <\gamma\leq 3$. We now introduce a random variable $K$, assigning the value of the degree of each vertex of the graph, i.e., $k$ is a particular value of $K$, implying $K$ ranges from $1$ to $N-1$. So, heterogeneity measurement tool -- the variance, and expectation, respectively, are: $var(K)=\sum_{k=1}^{N-1}k^{2}.p(k)-\sum_{k=1}^{N-1}k.p(k)=\sum_{k=1}^{N-1}(k^{2}-k).k^{-\gamma}$
and $z= \sum_{k=1}^{N-1}k.p(k)=\sum_{k=1}^{N-1}k.k^{-\gamma}$. Thus, $var(K)$ and $z$ can only vary with varying the value of $\gamma$, and the concept of the variance to be varied while expectation being constant is not realistic in any sense; moreover, both the two measures are either increasing or decreasing simultaneously.  Consequently, every viewed observation in this article can be read by the variance --  heterogeneity measurement tool of degree distribution.

Finally, we have reached to our prefixed destination, but this is not our final destination because the model runs throughout the assumption on unchanged population size over time. However, evolutionary dynamics and ecological process, those 
  are dependent on frequency dependent selection and demographic fluctuations, should be altered in fluctuation of population size (see \citep{huang2015stochastic}). In this connection, we also know that the adaptive network dynamics in which node states and network topologies dynamically change adaptively to each other is well-establish an area in computational network science where coevolution rules aim to integrate the adaptive network dynamics into the framework of evolutionary games -- for useful reviews see \citep{perc2010coevolutionary, perc2017statistical} and  for details see the references therein. Therefore, it is time to come to face the real great challenge to derive the analytical procedure to measure the evolution of cooperation in the fluctuation of population size; that will be a base model for the corresponding adaptive network simulation. In doing that we will drastically have to change the basic structure. Our surgical endeavours would surely lead us toward our ultimate destination; however, it will take time.

\section*{Acknowledgments}
I acknowledge the supporting advice of Constantino Tsallis and would like to thank  two anonymous reviewers for their valuable comments and constructive suggestions on the standard of the presentation and explanation of the manuscript.



\begin{thebibliography}{10}
\expandafter\ifx\csname url\endcsname\relax
  \def\url#1{\texttt{#1}}\fi
\expandafter\ifx\csname urlprefix\endcsname\relax\def\urlprefix{URL }\fi
\expandafter\ifx\csname href\endcsname\relax
  \def\href#1#2{#2} \def\path#1{#1}\fi

\bibitem{nowak2004emergence}
M.~A. Nowak, A.~Sasaki, C.~Taylor, D.~Fudenberg, Emergence of cooperation and
  evolutionary stability in finite populations, Nature 428~(6983) (2004)
  646--650.

\bibitem{ohtsuki2006simple}
H.~Ohtsuki, C.~Hauert, E.~Lieberman, M.~A. Nowak, A simple rule for the
  evolution of cooperation on graphs, Nature 441~(7092) (2006) 502.

\bibitem{taylor2007evolution}
P.~D. Taylor, T.~Day, G.~Wild, Evolution of cooperation in a finite homogeneous
  graph, Nature 447~(7143) (2007) 469.

\bibitem{zukewich2013consolidating}
J.~Zukewich, V.~Kurella, M.~Doebeli, C.~Hauert, Consolidating birth-death and
  death-birth processes in structured populations, PLoS One 8~(1) (2013)
  e54639.

\bibitem{taylor2004evolutionary}
C.~Taylor, D.~Fudenberg, A.~Sasaki, M.~A. Nowak, Evolutionary game dynamics in
  finite populations, Bulletin of Mathematical Biology 66~(6) (2004)
  1621--1644.

\bibitem{traulsen2005coevolutionary}
A.~Traulsen, J.~C. Claussen, C.~Hauert, Coevolutionary dynamics: from finite to
  infinite populations, Physical Review Letters 95~(23) (2005) 238701.

\bibitem{du2015aspiration}
J.~Du, B.~Wu, L.~Wang, Aspiration dynamics in structured population acts as if
  in a well-mixed one, Scientific Reports 5 (2015) 8014.

\bibitem{fletcher2009simple}
J.~A. Fletcher, M.~Doebeli, A simple and general explanation for the evolution
  of altruism, Proceedings of the Royal Society of London B: Biological
  Sciences 276~(1654) (2009) 13--19.

\bibitem{perc2013evolutionary}
M.~Perc, J.~G{\'o}mez-Garde{\~n}es, A.~Szolnoki, L.~M. Flor{\'\i}a, Y.~Moreno,
  Evolutionary dynamics of group interactions on structured populations: a
  review, Journal of the Royal Society Interface 10~(80) (2013) 20120997.

\bibitem{perc2011does}
M.~Perc, Does strong heterogeneity promote cooperation by group interactions?,
  New Journal of Physics 13~(12) (2011) 123027.

\bibitem{szabo2007evolutionary}
G.~Szab{\'o}, G.~Fath, Evolutionary games on graphs, Physics reports 446~(4)
  (2007) 97--216.

\bibitem{matsuda1992statistical}
H.~Matsuda, N.~Ogita, A.~Sasaki, K.~Sat{\=o}, Statistical mechanics of
  population: the lattice lotka-volterra model, Progress of Theoretical Physics
  88~(6) (1992) 1035--1049.

\bibitem{morris1997representing}
A.~J. Morris, Representing spatial interactions in simple ecological models,
  Ph.D. thesis, University of Warwick (1997).

\bibitem{van1998unit}
M.~Van~Baalen, D.~A. Rand, The unit of selection in viscous populations and the
  evolution of altruism, Journal of Theoretical Biology 193~(4) (1998)
  631--648.

\bibitem{house2011insights}
T.~House, M.~J. Keeling, Insights from unifying modern approximations to
  infections on networks, Journal of The Royal Society Interface 8~(54) (2011)
  67--73.

\bibitem{hadjichrysanthou2012approximating}
C.~Hadjichrysanthou, M.~Broom, I.~Z. Kiss, Approximating evolutionary dynamics
  on networks using a neighbourhood configuration model, Journal of Theoretical
  Biology 312 (2012) 13--21.

\bibitem{konno2011condition}
T.~Konno, A condition for cooperation in a game on complex networks, Journal of
  Theoretical Biology 269~(1) (2011) 224--233.

\bibitem{ohtsuki2006replicator}
H.~Ohtsuki, M.~A. Nowak, The replicator equation on graphs, Journal of
  Theoretical Biology 243~(1) (2006) 86--97.

\bibitem{santos2006evolutionary}
F.~C. Santos, J.~M. Pacheco, T.~Lenaerts, Evolutionary dynamics of social
  dilemmas in structured heterogeneous populations, Proceedings of the National
  Academy of Sciences of the United States of America 103~(9) (2006)
  3490--3494.

\bibitem{perc2008social}
M.~Perc, A.~Szolnoki, Social diversity and promotion of cooperation in the
  spatial prisoner’s dilemma game, Physical Review E 77~(1) (2008) 011904.

\bibitem{santos2012role}
F.~C. Santos, F.~L. Pinheiro, T.~Lenaerts, J.~M. Pacheco, The role of diversity
  in the evolution of cooperation, Journal of Theoretical Biology 299 (2012)
  88--96.

\bibitem{xu2015evolution}
B.~Xu, M.~Li, R.~Deng, The evolution of cooperation in spatial prisoner’s
  dilemma games with heterogeneous relationships, Physica A: Statistical
  Mechanics and its Applications 424 (2015) 168--175.

\bibitem{perc2009evolution}
M.~Perc, Evolution of cooperation on scale-free networks subject to error and
  attack, New Journal of Physics 11~(3) (2009) 033027.

\bibitem{szolnoki2016collective}
A.~Szolnoki, M.~Perc, Collective influence in evolutionary social dilemmas, EPL
  (Europhysics Letters) 113~(5) (2016) 58004.

\bibitem{perc2017statistical}
M.~Perc, J.~J. Jordan, D.~G. Rand, Z.~Wang, S.~Boccaletti, A.~Szolnoki,
  Statistical physics of human cooperation, Physics Reports 687 (2017) 1--51.

\bibitem{santos2008social}
F.~C. Santos, M.~D. Santos, J.~M. Pacheco, Social diversity promotes the
  emergence of cooperation in public goods games, Nature 454~(7201) (2008) 213.

\bibitem{szolnoki2008towards}
A.~Szolnoki, M.~Perc, Z.~Danku, Towards effective payoffs in the prisoner’s
  dilemma game on scale-free networks, Physica A: Statistical Mechanics and its
  Applications 387~(8) (2008) 2075--2082.

\bibitem{pinheiro2012local}
F.~L. Pinheiro, J.~M. Pacheco, F.~C. Santos, From local to global dilemmas in
  social networks, PloS One 7~(2) (2012) e32114.

\bibitem{li2013evolution}
C.~Li, B.~Zhang, R.~Cressman, Y.~Tao, Evolution of cooperation in a
  heterogeneous graph: Fixation probabilities under weak selection, PloS One
  8~(6) (2013) e66560.

\bibitem{nowak2006five}
M.~A. Nowak, Five rules for the evolution of cooperation, Science 314~(5805)
  (2006) 1560--1563.

\bibitem{kimura1964diffusion}
M.~Kimura, Diffusion models in population genetics, Journal of Applied
  Probability 1~(2) (1964) 177--232.

\bibitem{Kampen1992}
N.~G. Van~Kampen, {Stochastic Processes in Physics and Chemistry }, Elsevier,
  Amsterdam, 1992.

\bibitem{Gardiner2009}
C.~W. Gardiner, {Handbook of Stochastic Methods for Physics, Chemistry and the
  Natural Sciences}, Springer, 1997.

\bibitem{gillespie1976general}
D.~T. Gillespie, A general method for numerically simulating the stochastic
  time evolution of coupled chemical reactions, Journal of Computational
  Physics 22~(4) (1976) 403--434.

\bibitem{gillespie1977exact}
D.~T. Gillespie, Exact stochastic simulation of coupled chemical reactions, The
  Journal of Physical Chemistry 81~(25) (1977) 2340--2361.

\bibitem{Morita2008}
S.~{Morita}, {Extended pair approximation of evolutionary game on complex
  networks}, Progress of Theoretical Physics 119 (2008) 29--38.

\bibitem{ohtsuki2008evolutionary}
H.~Ohtsuki, M.~A. Nowak, Evolutionary stability on graphs, Journal of
  Theoretical Biology 251~(4) (2008) 698--707.

\bibitem{iyer2016evolution}
S.~Iyer, T.~Killingback, Evolution of cooperation in social dilemmas on complex
  networks, PLoS Computational Biology 12~(2) (2016) e1004779.

\bibitem{allen2017evolutionary}
B.~Allen, G.~Lippner, Y.-T. Chen, B.~Fotouhi, N.~Momeni, S.-T. Yau, M.~A.
  Nowak, Evolutionary dynamics on any population structure, Nature 544~(7649)
  (2017) 227--230.

\bibitem{maciejewski2014evolutionary}
W.~Maciejewski, F.~Fu, C.~Hauert, Evolutionary game dynamics in populations
  with heterogenous structures, PLoS Computational Biology 10~(4) (2014)
  e1003567.

\bibitem{huang2015stochastic}
W.~Huang, C.~Hauert, A.~Traulsen, Stochastic game dynamics under demographic
  fluctuations, Proceedings of the National Academy of Sciences 112~(29) (2015)
  9064--9069.

\bibitem{perc2010coevolutionary}
M.~Perc, A.~Szolnoki, Coevolutionary games -- a mini review, BioSystems 99~(2)
  (2010) 109--125.

\end{thebibliography}

\end{document}